\documentclass[aip,reprint,onecolumn]{revtex4-1}
\usepackage[dvips]{graphicx}
\usepackage{amssymb,amsfonts,amsmath}
\usepackage{color}

\newcommand{\bs}{{\bf {s}}}
\newcommand{\br}{{\bf {r}}}

\newcommand{\etal}{{\it et al.}}

\newcommand{\Vmy}[1]{\mathbf{#1}}
\newcommand{\rmi}{\mathrm{i}}
\newcommand{\rme}{\mathrm{e}}
\newcommand{\R}{\ensuremath\mathbb{R}}

\newcommand{\Sum}{\displaystyle \sum}
\newcommand{\Fra}[2]{\displaystyle \frac{ #1}{ #2}}

\newcommand{\Int}{\displaystyle \int}

\newcommand{\de}{\,\mathrm{d}}

\newcommand{\Dpar}[2]{\displaystyle \Fra{\partial #1}{\partial #2}}

\newcommand{\Dparxy}[3]{\displaystyle \Fra{\partial^2 #1}{\partial #2 \partial #3}}

\newcommand{\Lapl}[1]{\displaystyle \boldsymbol{\nabla}^2 #1}
\newcommand{\Div}[1]{\displaystyle \boldsymbol{\nabla} \cdot #1}

\newcommand{\Grad}[1]{\displaystyle \boldsymbol{\nabla} #1}

\begin{document}


\title{Quantum vortex reconnections} 



\author{S.~Zuccher}

\author{M.~Caliari}
\affiliation{Dipartimento di Informatica, Facolt\`a di Scienze,
Universit\`a di Verona, Ca' Vignal 2, Strada Le Grazie 15, 
37134 Verona, Italy}

\author{A.~W.~Baggaley}
\affiliation{School of Mathematics and Statistics, 
University of Glasgow, Glasgow, G12 8QW, United Kingdom}
\affiliation{Joint Quantum Centre (JQC) Durham-Newcastle, and
School of Mathematics and Statistics, 
Newcastle University, Newcastle upon Tyne, NE1 7RU, United Kingdom}
\author{C.~F.~Barenghi}
\affiliation{Joint Quantum Centre (JQC) Durham-Newcastle, and
School of Mathematics and Statistics, 
Newcastle University, Newcastle upon Tyne, NE1 7RU, United Kingdom}


\date{\today}

\begin{abstract}
We study reconnections of quantum vortices by numerically
solving the governing Gross-Pitaevskii equation. We find that the
minimum distance between vortices scales differently with time before
and after the vortex reconnection. 
We also compute vortex reconnections using 
the Biot-Savart law for vortex filaments of infinitesimal thickness, and
find that, in this model, reconnection are time-symmetric.
We argue that the likely cause of the difference between the Gross-Pitaevskii
model and the Biot-Savart model is the
intense rarefaction wave which is radiated away from a Gross-Pitaeveskii
reconnection. Finally we compare our results to experimental observations
in superfluid
helium, and discuss the different length scales probed by the two models and
by experiments.
\end{abstract}

\pacs{\\
47.32.C- (vortex dynamics)\\
67.30.he (vortices in superfluid helium)\\
03.75.Lm (vortices in Bose Einstein condensates)
}
\maketitle 

\section{Introduction}
\label{sec:intro}

The importance of vortex reconnections in turbulence~\cite{KT1994} 
cannot be understated.  Reconnections
randomize the velocity field, play a role in the
energy cascade, contribute to the fine-scale mixing, enhancing
diffusion~\cite{D2005,H1983,H1986}, and are the
dominant mechanism of jet noise generation.
If the axes of tubular vortex structures are interpreted as the skeleton
of turbulence, then the knottedness of the axes characterizes the 
turbulence's topology,
and vortex reconnections are the critical events which
change this topology~\cite{PSBS2003}. 
This idealized picture becomes physical reality if one moves from 
ordinary viscous fluids to quantum fluids~\cite{D1991}
such as superfluid liquid helium ($^3$He-B and $^4$He) and atomic Bose-Einstein
condensates. 
In these superfluid systems, quantum mechanics
constrains any rotational motion to vortex lines around which
the circulation is fixed by the condition

\begin{equation}
\oint_C \Vmy{u} \cdot \Vmy{dr}=\frac{h}{m}=\kappa,
\label{eq:circulation}
\end{equation}

\noindent
where $\Vmy{u}$ is the velocity,
$C$ is a closed integration path around the vortex axis, $h$ is Planck's
constant,  $m$ is the mass of the relevant boson (a helium atom in the case
of $^4$He, a Cooper pair in the case of $^3$He-B), and $\kappa$
is the quantum of circulation. Another constraint of quantum
mechanics is the small vortex core, which has fixed radius
($10^{-8}~\rm cm$ in $^4$He, $10^{-6}~\rm cm$ in $^3$He-B) and is
orders of magnitude smaller than the average distance between vortices in
typical experiments; because of these constraints,
no intensification or diminution of vorticity 
through stretching of the vortex core is possible in quantum fluids.

Vortex reconnections of individual quantum vortex lines are 
discrete, dramatic events, which have been recently visualized in the
laboratory~\cite{PFL2010}. They are the key to understanding 
quantum turbulence~\cite{SS2012}, a disordered state of
vortex lines which is easily created by
stirring liquid helium and atomic condensates.
Compared to quantum reconnections,
viscous reconnections in ordinary (classical) fluids are not
complete events: classical vorticity is continuous, not discrete,
and parts of the initial vortical tubes can be left behind as
vortex threads, which then undergo successive reconnections (the cascade and 
mixing scenarios), as newly formed vortex bridges recoil from each other by
self-advection~\cite{HD2011}. Another important difference is that
classical reconnections are dissipative events:
viscous forces turn
part of the fluid's energy into heat, whereas in a superfluid the
viscosity is zero, and the fluid's total energy is conserved.

It can be argued that, because of the utter simplicity of quantum
vortices (zero viscosity, fixed circulation, small
core size), quantum vortex reconnections are not only important 
phenomena of low temperature physics and of atomic physics,
but are also relevant to our general understanding of fluid phenomena
as toy models of Euler dynamics. 

Although a rather large number of studies addressed the viscous reconnection
problem, both the underlying mechanism and various scaling relationships have
remained elusive~\cite{KT1994}.
The first analytical work on classical vortex reconnections goes back to
Crow~\cite{C1970},
who studied the instability of a pair of counter-rotating vortex tubes 
shed from the wing tips of an airplane.
However, the systematic study of vortex reconnections began with
observations and laboratory experiments of the simplest and most
fundamental interaction of two colliding vortex
rings~\cite{FT1975,OA1977,OI1988}.
With the rapid development of supercomputers, direct numerical study of
vortex reconnection became possible.
Ashurst \& Meiron~\cite{AM1987} numerically solved the incompressible
Navier-Stokes equations in the region of closest approach of two vortex 
rings by providing an initial condition generated using the 
Biot-Savart (BS) model~\cite{Saffman} of vortex filaments.
At the same time, Pumir \& Kerr~\cite{PK1987} performed  numerical    
simulations of interacting vortex tubes.
Many other studies followed (for a review of the extensive work up to 1994
see the review of Kida \& Takaoka~\cite{KT1994}), but the literature on classical vortex reconnections seemed
to fade after~1994~\cite{MBG2001,CKL2003,KL2003,K2005,S2006}.
The recent work by Hussain \& Duraisamy~\cite{HD2011} renewed the interest in
the mechanics of viscous vortex reconnection.
Their study focused on the direct, high resolution
 numerical simulation of the incompressible Navier-Stokes
equations over a wide range of vortex Reynolds numbers, $\rm Re$,
for two perturbed anti-parallel vortex tubes.
They found that the minimum distance $\delta$ between 
the vortex tube centroids
scales as $\delta(t) \sim (t_0-t)^{3/4}$ \emph{before} the
reconnection and as $\delta(t) \sim (t-t_0)^2$ \emph{after} the
reconnection, where $t$ is time and $t_0$ is the instant of smallest separation
between the vortex centroids.

The literature concerned with quantum reconnections is more limited.
The possibility of reconnections was first raised by
Feynman~\cite{Feynman} in his pioneering work on the applications
of quantum mechanics to superfluid helium.  Schwarz~\cite{S1985,S1988}
realized that vortex reconnections are necessary to account for quantum
turbulence. He modelled quantum vortices as classical vortex
filaments and proposed the Local Induction
Approximation (LIA~\cite{Ricca96}) as a practical alternative to the 
exact (but CPU-intensive) Biot-Savart law for the numerical study of 
superfluid vortex dynamics.

A few years later, Koplik \& Levine~\cite{KL1993} performed the first 
numerical simulation of quantum vortex reconnections
by solving the Gross-Pitaevskii equation (GPE) which governs the motion
of a Bose-Einstein condensate and is used as a model of superfluid helium.
They found that if two vortices are  nearly anti-parallel when the
large-scale fluid motion brings them together, they reconnect, thus
confirming the conjecture of Feynman and Schwarz.

Soon after, using the vortex filament model of Schwarz,
de~Waele \& Aarts~\cite{WA1994} numerically integrated the
Biot-Savart equation for the ideal-fluid velocity field and claimed the
existence of a universal route to reconnection
for all kind of initial vortex-antivortex arrangements: their calculations
showed that, when vortices approach each other, they always form the same
pyramidal cusp. They measured the minimum distance between vortices, $\delta$,
during the approach to reconnection as a function of time $t$, and found that
\begin{equation}
\delta(t) \approx \sqrt{\frac{\kappa}{2 \pi}(t_0-t)},
\label{eq:aarts}
\end{equation}

\noindent
where $t_0$ is the time of reconnection. Equation~(\ref{eq:aarts})
is consistent with the
dimensional argument that, if the only relevant parameter in
reconnection dynamics is the quantum of circulation, then
\begin{equation}
\delta(t) =A (\kappa \vert t_0-t \vert )^{\alpha},
\label{eq:scaling}
\end{equation}

\noindent
where $\alpha=0.5$ and $A$ is a dimensionless constant of order unity.

Leadbeater \etal~\cite{Leadbeater01} used the GPE model to study 
reconnections of vortex rings launched against each other,
and discovered that a sound wave (in the form
of a short rarefaction pulse) is emitted at a reconnection event.
The wave turns part of the initial kinetic energy of
the vortices into acoustic energy which is radiated away. 
Acoustic energy is also created by vortex acceleration 
\cite{Vinen01,Leadbeater03}. The effects are
clearly important
to make sense of the observed decay of quantum turbulence
at very low temperatures~\cite{Vinen-Niemela}.

Individual quantum vortex reconnections were first observed by
Paoletti~\etal~\cite{PFL2010} 
by analyzing the trajectories of solid hydrogen tracers in superfluid $^4$He.
They verified that the scaling~(\ref{eq:scaling}) with $\alpha=0.5$
holds before and after the reconnections, that is to say quantum reconnections
are statistically time reversible, unlike classical reconnections~\cite{HD2011}.
They concluded that the expression
\begin{equation}
\delta(t) =A (\kappa \vert t_0-t \vert )^{1/2}(1+c \vert t_0-t \vert),
\label{eq:paoletti}
\end{equation}

\noindent
is the best fit to their data, with 
wide distribution of values centred around $A \approx 1.1$ (larger than
$A=(2 \pi)^{-1/2} \approx 0.4$
found by de~Waele \& Aarts\cite{WA1994}) and $c \approx 0$.

The universality of the route to reconnections was questioned by
Tebbs~\etal~\cite{TYB2011} who performed a
series of numerical simulations of quantum reconnections using the
GPE.  They reproduced the pyramidal shape of the vortex lines 
observed by de~Waele \& Aarts~\cite{WA1994} 
for the initial configuration used
by these authors, but did not observe the same shape for other
configurations. They also confirmed the scaling~(\ref{eq:scaling}), 
again in agreement with de~Waele \& Aarts ~\cite{WA1994}, but
did not measure the time evolution  of $\delta$ after the reconnection.

Kursa~\etal~\cite{KBL2011} employed LIA, Biot-Savart and GPE simulations,
and showed that a single reconnection of two almost anti-parallel
quantum vortices can lead to the
creation of a cascade of vortex rings, provided that the angle between the
vortices is sufficiently small.

Kerr~\cite{K2011}, by means of numerical simulation of the
GPE, investigated the reconnection between a pair of
perturbed anti-parallel quantum vortices.
He argued that kinetic energy is converted into interaction energy and
eventually local kinetic energy depletion that is similar to
energy decay in a classical fluid, even though the governing equations are
Hamiltonian and energy conserving.

The present study aims at characterizing vortex reconnection in quantum fluids
by performing direct numerical simulations of the three-dimensional
GPE in different vortex configurations.
Our goal is to extract the minimum distance $\delta$ between
vortices as a function of time,  both before and after the reconnection,
compare results against classical reconnections,  quantum reconnections computed
with the Biot-Savart law, and experiments in superfluid helium.

The plan of the paper is the following.
In section~\ref{sec:model} we introduce the governing GPE, cast it
in dimensionless form, and present the straight vortex solution.
In section~\ref{sec:results} we describe the initial condition of
our numerical calculations in terms of the initial angle $\beta$ between the
vortex lines, and present computed vortex reconnections, 
paying attention to the
minimum distance $\delta$ between vortices before and after the
reconnection, and the sound wave which is generated. We also perform
reconnections using the Biot-Savart model.
In section~\ref{sec:conclusion} we compare 
GPE reconnections and Biot-Savart reconnections with previous
work and experiments, and draw the conclusions. The numerical method
which we use to solve the GPE is described in the Appendix.

\section{Model}
\label{sec:model}

The governing equation for a weakly-interacting Bose-Einstein
condensate is the GPE \cite{Roberts-Berloff}

\begin{equation}\label{eq:DimNLSE}
\rmi \hbar \frac{\partial \psi}{\partial t} = -\Fra{\hbar^2}{2m} \Lapl \psi +
V_0|\psi|^2\psi-E_0 \psi,
\end{equation}

\noindent
where $\psi(\Vmy{x},t)$ the macroscopic single-particle
wave function for $N$ bosons of mass $m$ at position $\Vmy{x}$ and time $t$,
$V_0$ is the strength of the repulsive interaction between the
bosons, $E_0$ is the chemical potential (the energy increase upon adding
a boson), $\hbar=h/(2\pi)$, and the normalization condition is

\begin{equation}
\Int|\psi|^2\de \Vmy{x} = N.
\label{eq:norm}
\end{equation}

\noindent
To study atomic condensates rather than liquid helium, 
a term of the form $V_\mathrm{trap} \psi$, where $V_\mathrm{trap}$ is a suitable
trapping potential (usually harmonic), is added to the right hand
side of equation \ref{eq:DimNLSE}. Hereafter we shall not
consider such term, but limit our work to homogeneous condensates. 
When applying the GPE to superfluid helium we must remember
that helium is a liquid, not a weakly interacting gas,
so the GPE model is more qualitative than quantitative.
For example, the dispersion relation of 
small perturbations from the uniform solution 
$\psi_{\infty}=\sqrt{E_0/V_0}$ of equation~\ref{eq:DimNLSE}
is

\begin{equation}
\omega^2=\frac{\hbar^2 k^4}{4 m^2}+\frac{E_0}{m} k^2,
\label{eq:dispersion}
\end{equation}

\noindent
where $\omega$ is the angular velocity, $k$ is the wavenumber and 
$c=\sqrt{E_0/m}$ is the speed of sound. Note that for $k<<1$ we have
$\omega \approx c k$ (phonons), and for $k>>1$ we have
$\omega \approx \hbar k^2/(2 m)$
(free particles), without the
roton minimum which is characteristic of superfluid helium \cite{D1991}. 
Another shortcoming of the GPE is the vortex core: more sophisticated models of
the helium vortex core exist \cite{Reatto1997}, but are not practical
for the study of complex dynamics such as vortex reconnections.

By applying the transformation

\begin{equation}
\Vmy{x} \to \Fra{\hbar}{\sqrt{2mE_0}} \Vmy{x},\quad\quad
t \to \Fra{\hbar}{2E_0}t,\quad\quad
\psi \to \sqrt{\Fra{E_0}{V_0}} \psi,
\label{eq:units}
\end{equation}

\noindent
where $\psi_{\infty}=\sqrt{E_0/V_0}$ is the uniform solution at rest
in an infinite domain, we cast the GPE in the following dimensionless
form

\begin{equation}
\frac{\partial \psi}{\partial t} = \Fra \rmi 2 \Lapl \psi + \Fra \rmi 2(1 - |\psi|^2)\psi.
\label{eq:GPE}
\end{equation}

\noindent
The quantity $\zeta_0=\hbar/\sqrt{2 m E_0}$ is called the healing length
or coherence length. It is the typical length scale over which the
wave function bends, and therefore determines the vortex core radius
(see next section) and the thickness of any superfluid boundary layer in
the presence of a wall.

The numerical method to numericaly solve the GPE is 
described in the Appendix. Figure~\ref{fig:1}~(left) shows that during
the time evolution the Hamiltonian energy, defined by
\begin{equation}
E=\int dV \left(\frac{1}{2} \vert \nabla \psi \vert^2 
+\frac{1}{4} (1-\vert \psi \vert^2)^2\right)
\label{eq:E}
\end{equation}
is sufficiently well conserved: the numerical error is less than one part
in $10^7$.
The fluid dynamics interpretation of the GPE arises from
the Madelung transformation

\begin{equation}
\psi = \sqrt\rho \rme^{\rmi S},
\label{eq:Madelung}
\end{equation}

\noindent
which yields the following equations

\begin{equation}\label{eq:GPEinNScont}
\Dpar{\rho}{t}+\Dpar{(\rho u_j)}{x_j}=0,
\end{equation}

\begin{equation}\label{eq:GPEinNSmom}
\rho \left(\Dpar{u_i}{t}+u_j\Dpar{u_i}{u_j} \right) =
-\Dpar{p}{x_i} + \Dpar{\tau_{ij}}{x_j},\quad\quad i=1,2,3
\end{equation}

\noindent
(written in tensorial notation), where  density and
velocity are

\begin{equation}
\rho=|\psi|^2, \qquad \qquad \Vmy{u}=\Grad{S},
\end{equation}

\noindent
and

\begin{equation}
p=\Fra{\rho^2}{4}
\quad\text{and}\quad
\tau_{ij}=\Fra 1 4 \rho\Dparxy{\ln\rho}{x_i}{x_j},
\end{equation}

\noindent
are the pressure and the so-called quantum stress. It is easy to
verify that the quantum stress term at the right hand side of in equation 
(\ref{eq:GPEinNSmom}) is negligible compared to the pressure term at
length scales larger than unity (the coherence length). Since, as
we shall see, the vortex core radius is of the order of the coherence length,
we conclude that, at scales larger than the vortex core, the GPE, expressed by
equations (\ref{eq:GPEinNScont}) and (\ref{eq:GPEinNSmom}), reduces to
the classical (compressible) Euler equations. Incompressible Euler dynamics
is achieved in the further limit of small velocity (compared to the
sound speed $c=\sqrt{E_0/m}$) at constant density
(again, away from the vortex cores).

We seek a two-dimensional solution of equation~\eqref{eq:GPE} 
that represents a straight vortex centred at the origin.
It is well-known that the classical two-dimensional Euler vortex of
circulation $\Gamma$ has azimuthal velocity
$u_\theta = \Gamma/(2\pi r)$ where $r=\sqrt{x^2+y^2}$ is the radius
and $\theta=\arctan (y/x)$ is the azimuthal angle.
The Cartesian components of the velocity are thus
$u_x=-u_\theta \sin\theta = -\Gamma y/(2 \pi r^2)$ and 
$u_y=u_\theta \cos\theta =\Gamma x /r^2$.
Therefore $\Vmy{u}=(u_x,u_y)=(\Gamma/(2 \pi))\Grad{\theta}$. This shows
that the velocity field is solenoidal ($\Div{\Vmy{u}}=0$), that the
quantum mechanical phase, $S$, is simply the azimuthal angle $\theta$, and
that the quantum of circulation, in our dimensionless units, is
equal to $2 \pi$.  
In steady conditions, the continuity equation
ensures that $\Div{(\rho\Vmy{u})}=0$, hence
$\Vmy{u}\cdot\Grad{\rho}=0$, which means that
$ \Grad{\rho}\cdot \Grad{\theta} = 0$. The solution
$\rho = \mathrm{const}$ has infinite energy and must be rejected. The
other possibility is that $\Grad{\rho} \perp \Grad{\theta}$. Since the
cylindrical coordinates $r$ and $\theta$ are
perpendicular to each other, $\Grad{\theta}$ is parallel to $r$, hence
$\Grad{\rho}$ is perpendicular to $r$.  Therefore, for a steady
two-dimensional and divergence-free velocity field, we must have
$\rho=\rho(r)$. Seeking a solution which represents a vortex
centred at the origin, we set
$\psi=\sqrt\rho \rme^{\rmi S}= f(r)\rm e^{\rmi \theta}$ where $f(r)$ is
a function to be determined. By imposing
that $\psi$ is the steady solution of equation~\eqref{eq:GPE}, we find that
$f(r)$ satisfies the equation

\begin{equation}
f'' + \Fra{f'}{r} + f \left(1 - f^2 -\Fra{1}{r^2} \right)=0
\label{eq:odef}
\end{equation}

\noindent
with boundary conditions $f(0) = 0$, $f(\infty)=1$. The equation
can be integrated numerically, but, to make
the computation of the initial condition faster, we look for
a Pad\'e approximation $R(r)$ to $f(r)$ of the form
\begin{equation}
R(r)=
\frac{\sum_{j=0}^{m}p_j r^j}{1+\sum_{k=1}^{n}q_k r^k}=
\frac{p_0+p_1r+p_2r^2+\cdots+p_mr^m}{1+q_1 r+q_2r^2+\cdots+q_nr^n},
\end{equation}

\noindent
which agrees with $f(r)$ at the origin to the highest possible order,
i.e. $f(0)=R(0)$, $f'(0)=R'(0)$, $f''(0)=R''(0)$, $\dots$,
$f^{(m+n)}(0)=R^{(m+n)}(0)$.
In our case $\rho \ge 0$,  and the Pad\'e approximation can be limited
to~\cite{BER2004}

\begin{equation}
\rho(r) \approx \Fra{r^2(c_1 + c_2 r^2)}{1+c_3 r^2+c_2 r^4},
\end{equation}

\noindent
with
\begin{equation}
c_1=\Fra{11}{32}, \quad
c_3=\Fra{5-32 c_1}{48-192 c_1}, \quad
c_2=c_1\left( c_3-\frac 1 4 \right).
\end{equation}

\noindent
The quantum vortex core is thus a hole of radius of the order of $a_0$
around which the quantum mechanical phase changes by $2 \pi$. 
Figure~\ref{fig:1}~(right) compares the radial profile of the density near
the vortex axis computed numerically and with the Pad\'e approximation.

\section{Results}
\label{sec:results}

\subsection{GPE initial condition}
\label{sec:results_init}

Our initial condition consists of two straight vortex lines $\Vmy{v}_1$ and
$\Vmy{v}_2$ as in figure~\ref{fig:2}
(the arrow denotes the direction of the vorticity),
 which intersect the $y$-axis respectively at the points  $C_1$ and $C_2$.
We call $\pi_1$ the $y$-$z$ plane, $\pi_2$ the plane perpendicular to $\pi_1$
and passing through $C_2$, and $s$ the intersection between $\pi_1$ and
$\pi_2$.  We call $\beta$ the angle formed by the directions $\Vmy{v}'_1$ and
$\Vmy{v}_2$, where $\Vmy{v}'_1$ is the projection of $\Vmy{v}_1$ onto $\pi_2$.
In the case of two parallel vortices we have $\beta=0$; in the case
of two anti-parallel vortices we have $\beta=\pi$.
Perpendicular intersections occur for $\beta=\pi/2$ and
$\beta=-\pi/2$ (the latter is show at the right of figure~\ref{fig:2}).

\subsection{GPE reconnection of anti-parallel vortices}
\label{sec:results_antiparallel}

Our first numerical experiment is concerned with the reconnection
of anti-parallel vortex lines (reconnecting angle $\beta=\pi$).
The size of the computational box is $-30 \le x,y,z \le 30$.
At time $t=0$ the vortex lines are located at position
$(x_0;y_0)=(10;\pm 3)$.  In order to 
make sure that the
reconnection occurs in the centre of the computational domain,
we impose a slight initial perturbation to the vortex lines, of the form
$A \left[\cos (2 \pi(z_\mathrm{max} - z_\mathrm{min})/\lambda)\right]^6$.
Figure~(\ref{fig:3}) shows some snapshots of the time-evolution.
It is apparent that the two vortices move as a pair in the $x$-direction.
The slight initial curvature enhances the Crow instability and the
vortices approach each other. The reconnection results in the
formation of two U-shaped vortex lines, which bend and
move apart from each other.

Figure~(\ref{fig:4}) shows the minimum distance $\delta$ between
reconnecting vortices as a function of time $t$. The algorithm to
compute $\delta$ searches for vortex lines, starting from the
 boundaries, following the minimum of the density $\rho$.
Once the vortex lines are retrieved, they are put in parametric form as
curves in $\R^3$, i.e. $\Vmy{r}(\xi)=(x(\xi),y(\xi),z(\xi))$, and
$\delta$ is computed.
Our main finding is that $\delta(t)$ behaves differently before and after
the reconnection. In this respect, quantum vortex reconnections are
therefore similar to viscous reconnections. Indeed
figure~(\ref{fig:3}) resembles very closely
figure~(3b) of Hussain \& Duraisamy~\cite{HD2011}.

\subsection{GPE reconnection of vortices at different initial angles}
\label{sec:results_angles}

We perform several other numerical calculations of vortex
reconnections to check 
whether the time dependence of $\delta(t)$
depends on the angle $\beta$ of the initial condition.
Figure~(\ref{fig:5}) shows the time evolution of
two initially straight vortices set at
the angle $\beta=3 \pi/4$.
Again, we observe that the vortices first approach each other, then move away
after the reconnection.
A very similar behavior characterize vortices initially set at
$\beta=\pi/2$ (orthogonal reconnection), 
whose time-evolution is shown in figure~(\ref{fig:6}).

Figure~(\ref{fig:7}) summarizes the results obtained for different values
of $\beta$.
The minimum distance between the vortices, $\delta$, 
is reported separately before and
after reconnection as a function of $|t-t_0|$, where $t_0$ is the time at
which the vortex reconnection takes place.
Power laws (black solid lines) of the form 
\begin{equation}
\delta = A \vert t-t_0\vert^{\alpha},
\label{eq:ourscaling}
\end{equation}

\noindent
(where $\delta$ and $t-t_0$ are dimensionless)
are super-imposed to fit the numerical data; the fitting
coefficients $A$ and $\alpha$
are reported in table~\ref{tab:1}. 
From Figure~(\ref{fig:7}) and table~(\ref{tab:1}) we conclude
that GPE reconnections are not time-symmetric: the average values
of $\alpha$ is $\alpha=0.39$ before the reconnection and
$\alpha=0.68$ after the reconnection. 
If we average all values, before and after,
we obtain $\alpha=0.53$, which is in fair agreement with 
$\alpha=0.5$ predicted by the scaling argument (\ref{eq:scaling}).
Table~(\ref{tab:1}) also shows that the average values of $A$ are
$A=1.29$ before the reconnection and $A=1.54$ after the reconnection. 
If we set $\alpha=0.5$ and return to dimensional variables,
we obtain
$A=1.29/\sqrt{2 \pi}=0.52$ and $1.54/\sqrt{2 \pi}=0.62$, which 
are about half of the value $A \approx 1.1$ found by 
Paoletti \etal~\cite{PFL2010} in their experiment.

As a check, we repeat all numerical simulations for opposite
angles (e.g. $\beta = \pm \frac \pi 2$,  $\beta = \pm \frac 3 4 \pi$, etc.)
obtaining the same temporal dependence of $\delta$ upon time $t$.

\subsection{GPE reconnection wave}
\label{sec:results_wave}

Our calculation confirms the finding of Leadbeater \etal~\cite{Leadbeater01},
that a vortex reconnection creates a wave.
First we consider the reconnection of anti-parallel vortices ($\beta=\pi/2$).
By extracting the iso-surfaces at quite large level, as done in
figure~(\ref{fig:8}) for $\rho = 0.94$, one notices
the formation of a mushroom-shaped pressure (density) rarefaction
wave generated by the
reconnection. The wave becomes shallower as it moves away from the vortices.
The bottom plot of figure~(\ref{fig:8}) shows the vortex lines together with
the pressure wave obtained by replacing the vortex tubes from the isosurface at
$\rho = 0.94$ with the isosurface at $\rho = 0.2$.
The footprint of the wave is particularly visible in contours of
$\rho$ on the plane $y=0$, as shown in figure~(\ref{fig:9}).

The pressure wave
is clearly visible for relatively small angles between vortices, 
$\beta< \pi/2$, whereas for larger angles it becomes difficult to 
clearly track it and visualize it.
Figure~(\ref{fig:10}) shows the time evolution of the mushroom-shaped pressure
wave ejected after reconnection for $\beta=7 \pi/ 8$ (isosurfaces at
$\rho=0.94$).

\subsection{GPE reconnections and vortex rings}
\label{sec:results_rings}

As we have mentioned in Section~\ref{sec:intro}, secondary generation of vortex
rings following a reconnection event was observed in the numerical
simulations of 
Kursa et al.~\cite{KBL2011} and Kerr \cite{K2011}. Kursa et al. also
studied how the emission of vortex rings depends on the initial angle
between the vortices (almost antiparallel configurations favour the
generation of vortex rings following the Crow instability). 
We do not investigate further the generation
of vortex rings, since it was already studied in detail in cited
works. We only remark that if we make our computational box longer
in the $z$ direction, vortex rings generation becomes visible,
as shown in figure~\ref{fig:11}. As for the physical significance
of this effect, we notice that, according to a recent study of
Baggaley et al. \cite{Baggaley-Sherwin}, the distribution of
reconnecting angles $\beta$ depends on the nature of the quantum turbulence.
Quantum turbulence generated with grids or propellers seems classical
in nature (for example the kinetic energy is distributed on the length
scales according to the classical Kolmogorov $k^{-5/3}$ law where
$k$ is the wavenumber), and
contains coherent bundles of vortices which induce reconnections
at small angles $\beta$. On the contrary, quantum turbulence generated
thermally (e.g. counterflow turbulence) is spatially more random, 
and reconnections tend to be antiparallel ($\beta \approx \pi$).

\subsection{Biot-Savart reconnections}
\label{sec:results_BS}

It is instructive to compare reconnections computed with the GPE
with reconnections computed with the Biot-Savart law. The latter,
which is widely used to study quantum turbulence,
approximates vortex lines as space curves $\bs=\bs(\xi,t)$ of
infinitesimal thickness which move according to 

\begin{equation}
\frac{d \bs}{dt}=
-\frac{\kappa}{4 \pi} \oint_{\cal L} \frac{(\bs-\br) }
{\vert \bs - \br \vert^3}
\times {\bf d}\br,
\label{eq:BS}
\end{equation}

\noindent
where $\xi$ is arc length and
the line integral extends over the entire vortex configuration. 
Equation~(\ref{eq:BS}) expresses incompressible Euler dynamics in
integral form~\cite{Saffman}.
Physically, it assumes that the density of the fluid
is constant (zero Mach number limit) and that the vortex core is
much smaller than any other length scale in the flow 
(a small parameter must be introduced to de-singularise the
integral).  Vortex reconnections are forbidden by Euler
dynamics, therefore, when applying equation~(\ref{eq:BS}) to superfluid
helium, we must supplement it with an algorithmic reconnection
procedure which changes the topology
of two vortex filaments when the distance between them is less than
a prescribed cutoff value, as first explained by Schwarz~\cite{S1985,S1988}. 
The numerical techniques which we use to compute Biot-Savart evolution
are described in our previous papers
\cite{Baggaley-cascade,Baggaley-fluctuations}. Here it suffices to
say that the vortex filaments are discretized
into a variable number of points $\bs_j$  ($j=1, \cdots N$),
holding their relative distance approximately 
between $\Delta \xi$ and $\Delta \xi/2$
where $\Delta \xi$ represents the prescribed numerical resolution.
The reconnection algorithm which we use, which
is triggered when the vortex separation is closer than $\Delta \xi/2$,
has been already described in detail~\cite{Baggaley-reconnections}
and compared
to other algorithms used in the liquid helium literature. 
It must be stressed that, unlike some of our recent work on quantum turbulence
\cite{Baggaley-structures}, the results which we present here
do not use a tree-algorithm~\cite{Baggaley-tree}
to approximate and speed up the calculation of Biot-Savart integrals.

We perform our calculations in an open domain
with numerical resolution $\Delta \xi=0.0005~\rm cm$.
The initial condition of the first numerical calculation which we present
consists of two vortex rings of the same polarity and
radius $R=0.0477~\rm cm$ set parallel to each other,
side-by-side on the $xz$-plane and travelling
in the $y$-direction, initially at
distance $\Delta x=0.002~\rm cm$ from each other. 
The initial number of discretization points for the two rings is $N=1600$. 
The same initial condition was used by de Waele \& Aarts~\cite{WA1994}.
Figure~(\ref{fig:12}) shows the time evolution. Note that the
resulting vortex reconnection is locally anti-parallel ($\beta = \pi$). 
The initial condition
of the second numerical calculation consists of the same two rings, but
initially set perpendicular to each other. The time evolution is shown
in figure~(\ref{fig:13}). Note that the
resulting reconnection is locally orthogonal ($\beta=\pi /2$).

The minimum distance between vortices, $\delta$, for both parallel
and perpendicular rings, is shown in figure~(\ref{fig:14}). 
It is apparent that the temporal scaling is time symmetric
before and after the reconnection, in agreement with 
equation~(\ref{eq:scaling}) with $\alpha=0.5$
and with the results of de~Waele \& Aarts~\cite{WA1994}.  
The time-symmetry of Biot-Savart reconnections contrast the
time-asymmetry of GPE reconnections showed in the previous sections
and of classical reconnections~\cite{HD2011}. We tested the dependence of this result 
on the reconnection algorithm used~\cite{Baggaley-reconnections}, and found no difference 
in the scaling. This is perhaps not surprising as
a change of the reconnection algorithm would imply a change of $\delta$
of the order of $\Delta \xi \approx 10^{-4}~\rm cm$ only, whilst we measure
the evolution of $\delta$ up to distances of the order $10^{-2}~\rm cm$.

The coefficient $A$ however is not the same in all cases
(although, for the approach of parallel rings, it is in fair 
agreement with de~Waele \& Aarts~\cite{WA1994}). A similar spread was
observed by Tsubota \& Adachi~\cite{TA2011}.
The speed at which the vortex lines move away from each other
after the reconnection is faster than the speed at which they approach 
each other; this effect is also visible in figures~(\ref{fig:14}),
and qualitatively consistent with the findings obtained
with the GPE, see figure~(\ref{fig:7}).

\section{Conclusion}
\label{sec:conclusion}

Hussain \& Duraisamy~\cite{HD2011} have shown that, in
ordinary incompressible viscous fluids, the minimum separation $\delta$
between reconnecting vortex tubes behaves differently before
($\delta(t) \sim (t_0-t)^{3/4}$) and after ($\delta(t) \sim (t-t_0)^2$)
the reconnection at $t=t_0$. By solving the GPE we find
a similar time asymmetry, although with different power laws: 
$\delta(t) \sim (t_0-t)^{0.4}$ and $\delta(t) \sim (t-t_0)^{0.7}$ 
respectively, independently of the initial angle between the vortex
lines. On the contrary, by solving the Biot-Savart equation, 
we find that the scaling is time symmetric, 
with  $\delta (t) \sim \vert t_0-t \vert^{1/2}$ 
for both $t<t_0$ and $t>t_0$. 

What causes the difference? The main difference between the GPE model
and the Biot-Savart model is that the former is compressible and the
latter is not.  Clearly the rarefaction wave which
is generated at the GPE reconnection breaks the time symmetry,
transforming~\cite{Leadbeater01} some of the kinetic energy of the vortices 
into acoustic energy
which is radiated to infinity (in analogy with the viscous dissipation of
kinetic energy at Navier-Stokes reconnections). 
The fact that GPE
reconnections do not follow the power law (\ref{eq:scaling})
predicted by the simple dimensional argument is not surprising: at the
very small scales explored by solving the GPE, other parameters besides 
the quantum of circulation
may be relevant in determing $\delta$, for example the coherence length.

It is also clear that
the Biot-Savart model and the GPE model probe different length scales.
As remarked 
in Section~\ref{sec:model}, the GPE should converge to
incompressible Euler behaviour in the limit $v/c <<1$
of small velocity $v$ compared to the speed of sound $c$. The velocity of
a vortex strand of local curvature $R$ is approximately
$v \approx \kappa/R$, hence we expect to recover
the Biot-Savart law for $R >> \kappa/c$. However, as noticed earlier,
the GPE is only a qualitative model of helium. In estimating $\kappa/c$
we should not use the observed value $c=238~\rm m/s$, but rather
the value of $c$ which arises from the GPE itself, consistently
with the vortex core size resulting from the GPE. 
The first step is to identify the coherence length $\zeta_0$.
Experiments with ions and vortex rings by Rayfield and Reif 
suggest \cite{RR1964,BDV1983} that the radius of the vortex core is
$a_0=1.3 \times 10^{-8}~\rm cm$. Figure~\ref{fig:1}~(right) shows that  
the density raises from zero to half of its value 
at infinity at distance $1.5 \zeta_0$ (1.5 dimensionless units)
from the axis of the vortex. 
Taking this distance as the (arbitrary but reasonable) value of the vortex
core radius $a_0$, we have $\zeta_0=0.87 \times 10^{-8}~\rm cm$. 
The half-size of our computational box (30 coherence length units) thus
corresponds to $26 \times 10^{-8}~\rm cm$.
The second step is to find the sound speed in the GPE evolution.
From $c=\sqrt{E_0/m}$ and $\zeta_0=\hbar/\sqrt{2 m E_0}$, we have
$c=\hbar/(\sqrt{2} m \zeta_0)=129~\rm m/s$,
 where $m=6.64 \times 10^{-24}~\rm g$
is the mass of one helium atom. We conclude that GPE evolution
should become similar to Biot-Savart evolution for 
$R>>\kappa/c\approx 8 \times 10^{-8}~\rm cm$, that is to say for
$R>> 9 \zeta_0$ (radius of curvature much greater than 9 dimensionless units).
It is apparent in figures~\ref{fig:3},~\ref{fig:5} and~\ref{fig:6} that
the radius of curvature of the vortex line(s) near the reconnecting point,
which determines $\delta(t)$,
is still too small to satisfy this condition.

The Biot-Savart model assumes scales much larger 
than the vortex core, which is effectively neglected. So it is not
possible, as a matter of principle, to use the Biot-Savart model to study
behaviour at the scale of the coherence length. Neither is possible, for
practical computing reasons, to solve the GPE at the large scales
explored by Biot-Savart calculations such as those in
section~(\ref{sec:results_BS}).

The experimental observations of Paoletti \etal~\cite{PFL2010}
agree with the Biot-Savart results in terms of the time symmetry and the
exponent $\alpha=0.5$ of the
power law (but it must be noticed that, for the GPE model,
the average of the exponents before and after the reconnection
is the same $\alpha \approx 0.5$ found in the experiment).
In the experiment, the motion of the vortex lines 
was detected using solid hydrogen tracer particles of radius 
$R \approx 10^{-4}~\rm cm$, ten thousand times larger than the coherence
length in superfluid $^4$He. 
The rarefaction wave generated by GPE reconnections has a
wavelength of about ten times the coherence length; although it is
very deep initially, it quickly spreads out and vanishes as it moves away.
On the scale of the tracer particles, the density
of the fluid is thus constant, so it is not surprising that the
Biot-Savart model is a better approximation to the observed
dynamics of the vortex lines.

Finally, the spread of the values of the coefficient $A$ which we
compute and the similar spread observed in the experiment is likely to
arise from differences in the initial condition, geometry of nearby
vortex lines and the velocity gradients which they induce, as
discussed by Paoletti \etal~\cite{PFL2010}.

\begin{acknowledgments}
We thank the Leverhulme Trust and EPSRC for financial support.
\end{acknowledgments}

\appendix
\section{Numerical method for the GPE}
\label{sec:numerics}

Without loss of generality, and for sake of simplicity, we describe in detail
the numerical method applied to the one-dimensional case of
equation~\eqref{eq:GPE} and report at the end of the section the
straightforward generalization to the three-dimensional case.
It is convenient to split the GPE evolution in two time steps~\cite{BJM2003}:

\begin{equation}
\label{eq:splitone}
\psi_t = \Fra \rmi 2 \Lapl \psi,
\end{equation}
\begin{equation}
\label{eq:splittwo}
\psi_t = \Fra \rmi 2(1 - |\psi|^2)\psi,
\end{equation}

\noindent
thus separating
linear and non-linear operators, where the subscript $t$ denotes the
time derivative.
We assume that the solution is periodic in the domain $a \le x < b$, i.e.
$\psi(a)=\psi(b)$, and seek a numerical solution in the time
interval $0 \le t \le T$ by expanding $\psi$ via
Fourier transform as

\begin{equation}
\psi(x,t) = \Sum_{j=-\frac M 2}^{\frac M 2-1} \phi_j(t)\mathcal{F}_j(x),
\quad a \le x < b, \quad 0 \le t \le T,
\end{equation}

\noindent
where $M$ is the number of modes, $\phi_j(t)$ are the time-dependent Fourier
coefficients, and the functions
$\mathcal{F}_j(x)= \frac{1}{\sqrt{b-a}} \rme^{\rmi 2 \pi j \frac{x-a}{b-a}}$
are orthonormal (i.e.
$\int_a^b\mathcal{F}_j\overline{\mathcal{F}}_k \de x=
\int_a^b \mathcal{F}_j \mathcal{F}_{-k} \de x =\delta_{jk}$).
By computing the temporal and spatial derivatives of $\psi$,
substituting them in equation~(\ref{eq:splittwo}),
multiplying the latter times $\overline{\mathcal{F}}_k$,
integrating between $a$ and $b$, and using the orthonormality property
of the functions $\mathcal{F}$,
equation~(\ref{eq:splittwo}) reads
\begin{equation}
\phi_k '(t) = \Fra \rmi 2 \lambda_k \phi_k(t),
\quad - \Fra M 2 \le k <  \Fra M 2,\quad 0 \le t \le T,
\end{equation}

\noindent
where $\lambda_j=-\left(\frac{ 2 \pi j }{b-a}\right)^2$ is real and negative.
The solution is, trivially,

\begin{equation}
\phi_k(t) = \rme^{t \frac \rmi 2 \lambda_k } \phi_k(0).
\end{equation}

\noindent
Since both $t$ and $\lambda$ are real,
$| \phi_j (t) |^2 = | \phi_j (0) |^2$
for all $j$.
Therefore, the total mass $m$ is preserved:

\begin{equation}
m(t) =\| \psi (t) \|_{L^2}^2= \Int_a^b\psi (x,t) \overline{ \psi}(x,t) \de x =
\Sum_{j=-\frac M 2}^{\frac M 2-1} | \phi_j (t) |^2 =
\Sum_{j=-\frac M 2}^{\frac M 2-1} | \phi_j (t_0) |^2 =
\| \psi (0) \|_{L^2}^2= m(0).
\end{equation}

\noindent
Moreover, if $\vec\phi$ denotes the vector of Fourier coefficients
$\vec\phi = [\phi_{-\frac M 2},\dots,\phi_{\frac M 2}]$,
then

\begin{equation}
\| \psi (t) \|_{L^2}^2= \|\vec\phi(t)\|_2^2=\|\vec\phi(t_0)\|_2^2=m(t_0)=m,
\end{equation}

\noindent
and the total mass $m$ can be retrieved simply as
the square of the norm of the
complex-coefficient vector  $\vec\phi$.

The second part of the time-splitting scheme
preserves mass as well.
This is easy proved by taking the conjugate of equation~(\ref{eq:splittwo}); 
from $\psi_t = ({\rm i}/2)(1 - |\psi|^2)\psi$ we obtain
$\overline\psi_t = (-{\rm i}/2) (1 - |\psi|^2)\overline\psi$,
hence the derivative of $|\psi|^2$ with respect to $t$ is

\begin{equation}
\Dpar{|\psi|^2}{t}= \Dpar{\psi\overline\psi}{t}=
\psi_t\overline\psi + \psi \overline\psi_t =
\Fra \rmi 2(1 - |\psi|^2)\psi\overline\psi +
\psi\left[ -\Fra \rmi 2(1 - |\psi|^2)\overline\psi\right] = 0.
\end{equation}

\noindent
The time independence of $|\psi|^2$ is crucial because it implies that the
solution of equation~(\ref{eq:splittwo}) is simply

\begin{equation}
\psi(x,t)= \rme^{t \frac \rmi 2(1 - |\psi(x,0)|^2)}\psi(x,0).
\end{equation}

The above method can be generalised naturally to the
three-dimensional case.
The unknown function $\psi(x,y,z,t)$ is expanded as

\begin{equation}
\psi(x,y,z,t) =
\Sum_{j,k,l}\phi_{jkl}(t)\mathcal{F}_j(x) \mathcal{F}_k(y) \mathcal{F}_l(z),
\end{equation}

\noindent
where we use the notation 
$\sum_{j,k,l}=\sum_{j=-\frac{M_x}{2}}^{\frac{M_x}{2}}
\sum_{k=-\frac{M_y}{2}}^{\frac{M_y}{2}}
\sum_{l=-\frac{M_z}{2}}^{\frac{M_z}{2}}$.

After computing the temporal and spatial derivatives of $\psi$,
substituting them in equation~(\ref{eq:splittwo}),
multiplying the differential equation times 
$\overline{\mathcal{F}}_m(x) \overline{\mathcal{F}}_n(y)
\overline{\mathcal{F}}_s(z)$,
integrating in space, and using the orthonormality property
of the functions $\mathcal{F}$, the first part of the splitting
now becomes

\begin{equation}
\phi_{jkl}'(t) = \frac \rmi 2 \lambda_{jkl} \phi_{jkl}(t),
\end{equation}

\noindent
where
\begin{displaymath}
\lambda_{jkl}=-\left(\Fra{ 2 \pi j }{b_x-a_x}\right)^2
-\left(\Fra{ 2 \pi k }{b_y-a_y}\right)^2-\left(\Fra{ 2 \pi l }{b_z-a_z}\right)^2
\end{displaymath}

\noindent
is real and negative and the mass-preserving solution is

\begin{displaymath}
\phi_{jkl}(t) =\rme^{t \frac \rmi 2 \lambda_{jkl}}\phi_{jkl}(0).
\end{displaymath}

As explained earlier, the second part of the splitting preserves mass
too, and its solution is explicit.

In order to outline the numerical algorithm it is convenient to formulate the
partial differential equation~\eqref{eq:GPE} as an ordinary differential
equation suppressing the spatial dependence and replacing
$\psi(\cdot,t)$ with $u(t)$~\cite{CNT2009}.  
We obtain the initial value problem
\begin{equation}
u'(t) = \left[A + B(u(t)) \right] u(t), \quad u(t_0)=u_0,
\end{equation}

\noindent
Using second-order Strang splitting~\cite{S1968},
the solution $u(t)=u(k \Delta t)=u_k$ can be recursively determined by
the multiplication

\begin{equation}
u^{k+1}=\rme^{\frac{\Delta t}{2} B}\rme^{\Delta t A}
\rme^{\frac{\Delta t}{2} B} u^m.
\end{equation}

\noindent
In our case $A=\Fra \rmi 2 \Lapl $ and $B(u(t))=\Fra \rmi 2(1 - |u(t)|^2)$.
As previously shown, $|u(t)|^2=|\psi|^2$ is constant, thus both operators
$A$ and $B$ are linear.

Since both solutions of the two parts of the splitting are explicit,
the only numerical error introduced by the method is confined to the
computation of the Fourier transform and its inverse.
The second order error in time due to the Strang splitting can be improved 
to fourth order without further numerical complications~\cite{Y1990}.

In conclusion,
assuming that the initial condition $\psi(x,y,z,t_0)$ is periodic
in all  spatial directions, the resulting algorithm is:

\begin{enumerate}
\item $\widetilde{\psi} (x,y,z,t_0)=\rme^{\frac{\Delta t}{2} \frac \rmi
           2(1 - |\psi(t_0,x)|^2)}\psi(x,y,z,t_0)$: 
           operator $B$ is applied in physical space
\item $\vec \phi(t_0) = \mathrm{FFT}(\widetilde{\psi} (x,y,z,t_0))$:
           Fourier transform is applied
\item $\phi_{jkl}(t_0+\frac{\Delta t}{2}) =
\rme^{\Delta t \frac \rmi 2 \lambda_{jkl}}\phi_{jkl}(t_0) \; \forall \,i,j,k$:
operator $A=\Fra \rmi 2 \Lapl$ is applied in  Fourier space
\item $\widetilde{\psi} (x,y,z,t_0+\frac{\Delta t}{2})=
\mathrm{IFFT}(\vec \phi(t_0+\frac{\Delta t}{2}))$: inverse Fourier transform
is applied to go back to physical space
\item $\psi (x,y,z,t_0 + \Delta t)=\rme^{\frac{\Delta t}{2} \frac \rmi
           2(1 - |\widetilde{\psi} (x,y,z,t_0+\frac{\Delta t}{2})|^2)}
           \widetilde{\psi} (x,y,z,t_0+\frac{\Delta t}{2})$:
           operator $B=\Fra \rmi 2(1 - |u(t)|^2)$ is applied in physical space
\end{enumerate}

\noindent
Clearly, step~5 is needed only to retrieve the physical solution
$\psi_k(x,y,z)$ at a certain time $t=k\Delta t$, otherwise it
can be avoided by merging step~5
and step~1 a single time step $\Delta t$.

\noindent
The limit of this scheme is that the initial condition $\psi_0(x,y,z)$
must be periodic.
If it is not, it must be made periodic by adding image vortices on a
larger domain, i.e. more computational effort is required.
However, certain geometries (such as two anti-parallel vortices aligned along
the $z$-direction, moving along $x$ and centered in $(x_0;\pm y_0)$)
allow us to double the grid points only in
the $x$-direction due to the symmetry with respect to the $y=0$ and $z=0$ 
planes.
In general, if all vortices are aligned along one axis (typically $z$), image
vortices must be introduced in both other directions.
For general geometries images vortices must be introduced in all directions,
causing a memory allocation eight times larger.


\newpage


\begin{table}[h!!!]
\begin{center}
\begin{tabular}{lllll}
\hline
angle $\beta$ & $A_\mathrm{before}$ & $\alpha_\mathrm{before}$& $A_\mathrm{after}$ &
$\alpha_\mathrm{after}$\\
\hline
$\frac 8 8 \pi=\pi$  &1.36 &0.30 &1.88 & 0.66 \\
$\frac 7 8 \pi=\frac 7 8 \pi$ &1.23 & 0.39 & 2.71 & 0.63\\
$\frac 6 8 \pi=\frac 3 4 \pi$ &1.44 & 0.41 & 1.69 & 0.68\\
$\frac 5 8 \pi=\frac 5 8 \pi$ &1.35 & 0.44 & 1.30& 0.69\\
$\frac 4 8 \pi=\frac \pi 2$ &1.41 & 0.36 & 1.01 & 0.67\\
$\frac 3 8 \pi=\frac 3 8 \pi$ &0.94 & 0.42 & 0.66 & 0.73\\
\hline
average & 1.29 & 0.39 & 1.54 & 0.68\\
\hline
\end{tabular}
\end{center}
\caption{Coefficients of the fit $\delta(t)=A \vert t-t_0 \vert^{\alpha}$
of the minimum distance between vortices before and after the reconnection
at different initial angles $\beta$.}
\label{tab:1}
\end{table}

\newpage

\begin{figure}[h!!!]
\begin{center}
\includegraphics[scale=.85]{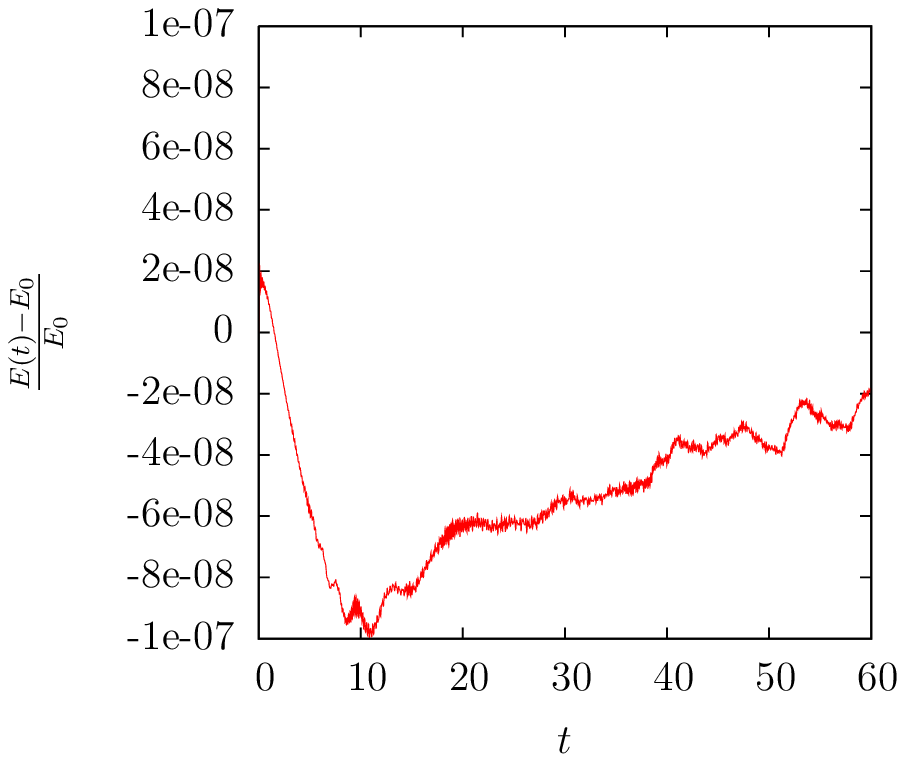} 
\includegraphics[scale=.85]{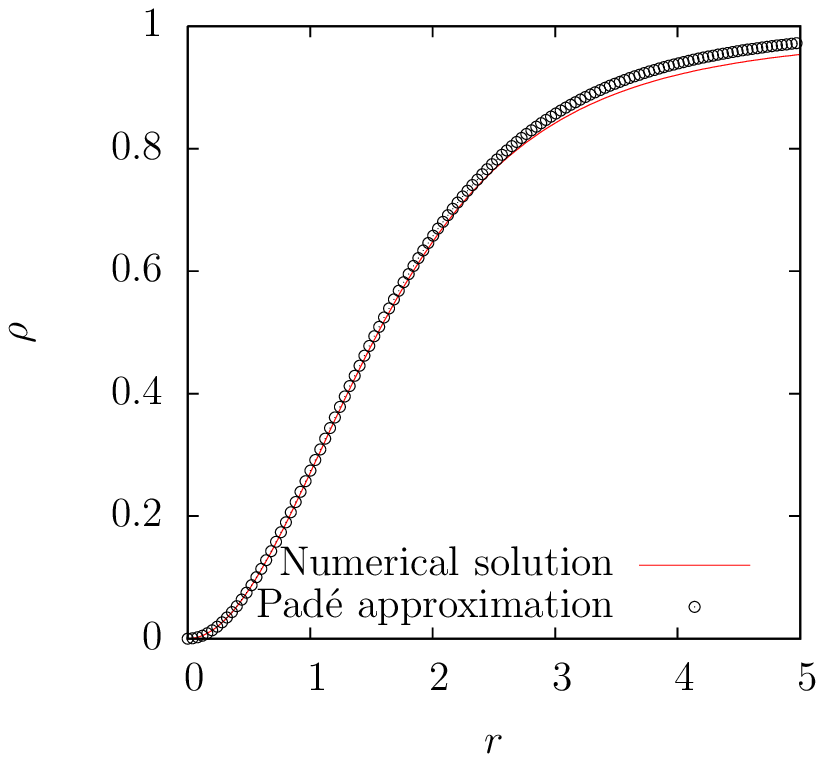} 
\end{center}
\caption{
Left: Relative change of the dimensionless
Hamiltonian energy $E(t)$ as a
function of dimensionless time $t$ 
with respect to the initial energy $E_0=E(0)$
during antiparallel reconnection.
Right:
Dimensionless density $\rho$ as a function of the
dimensionless radial coordinate $r$ computed numerically (solid line,
obtained by setting $r_{\infty}=20$ and using 500 grid points)
and with the Pad\'e approximation (empty circles).
}
\label{fig:1}
\end{figure}
 
\newpage

\begin{figure}[h!!!]
\begin{center}
\begin{minipage}{0.49\textwidth}
\makebox[\textwidth]{\includegraphics[scale=.35]{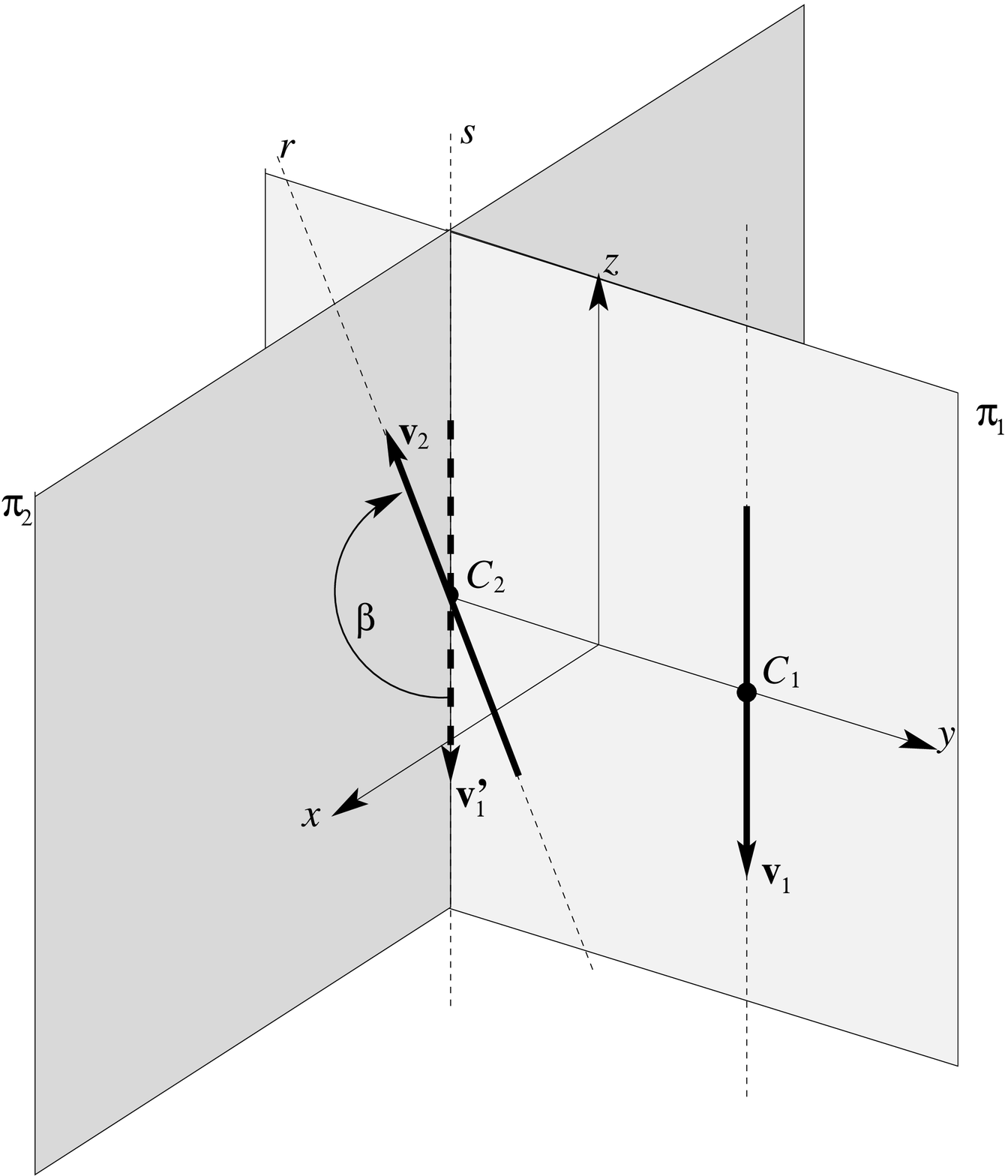}}
\end{minipage}
\begin{minipage}{0.49\textwidth}
\makebox[\textwidth]{\includegraphics[scale=.35]{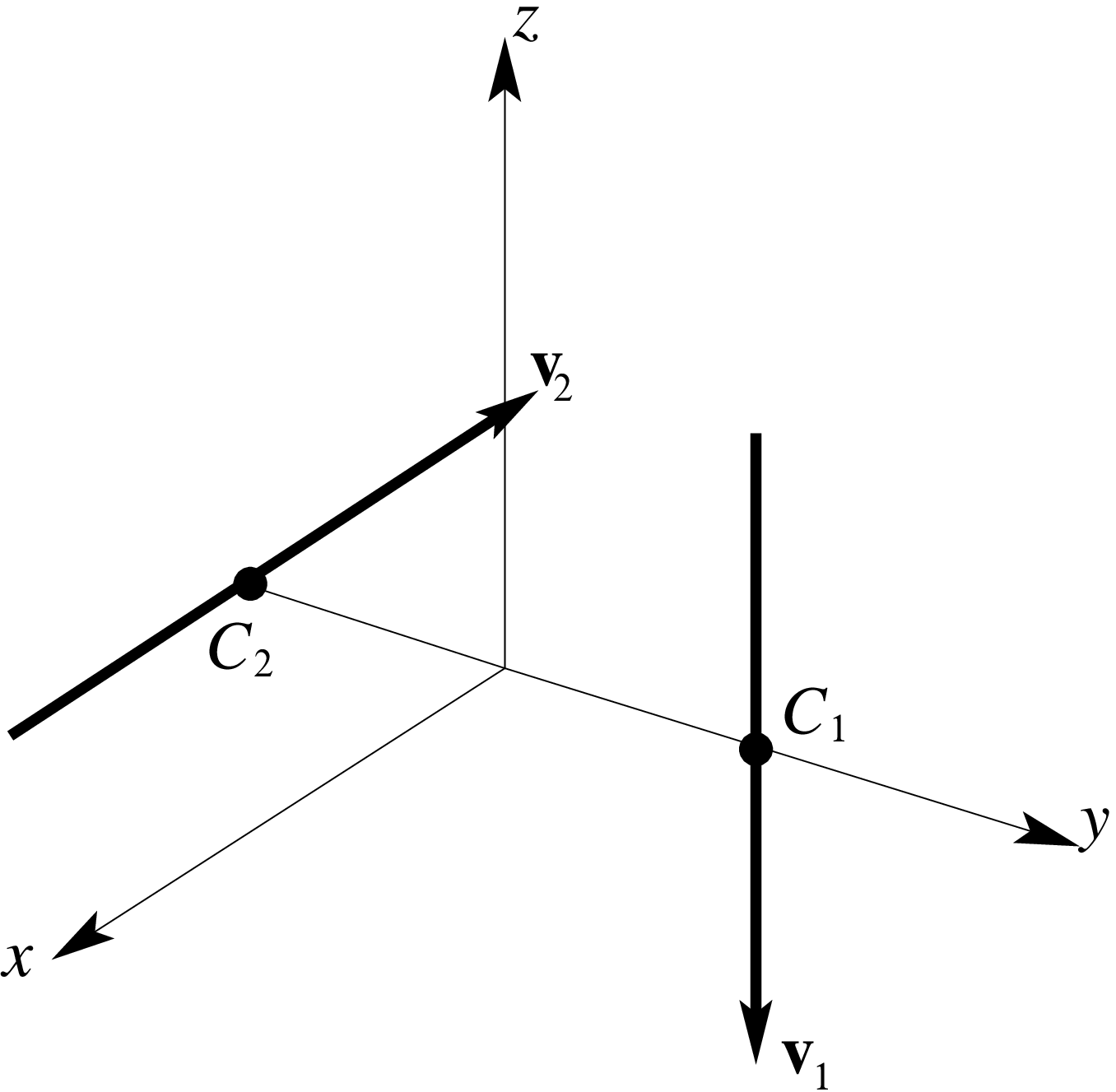}}
\end{minipage}
\end{center}
\caption{Initial condition. Left: $\beta=3 \pi/4$. Right: $\beta=-\pi/2$.}
\label{fig:2}
\end{figure}

\newpage

\begin{figure}[h!!!]
\begin{center}
\includegraphics[scale=.25]{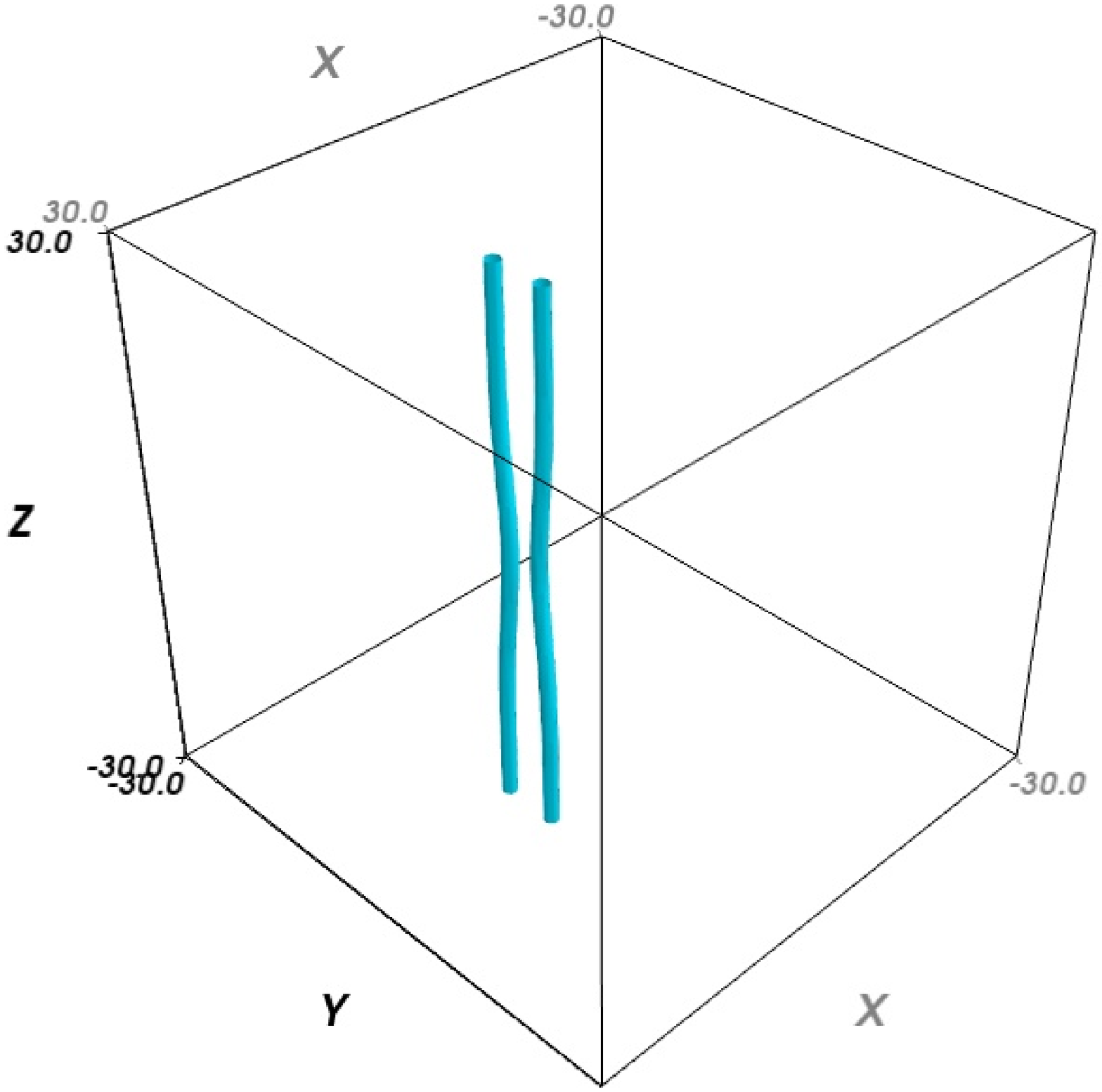} 
\includegraphics[scale=.25]{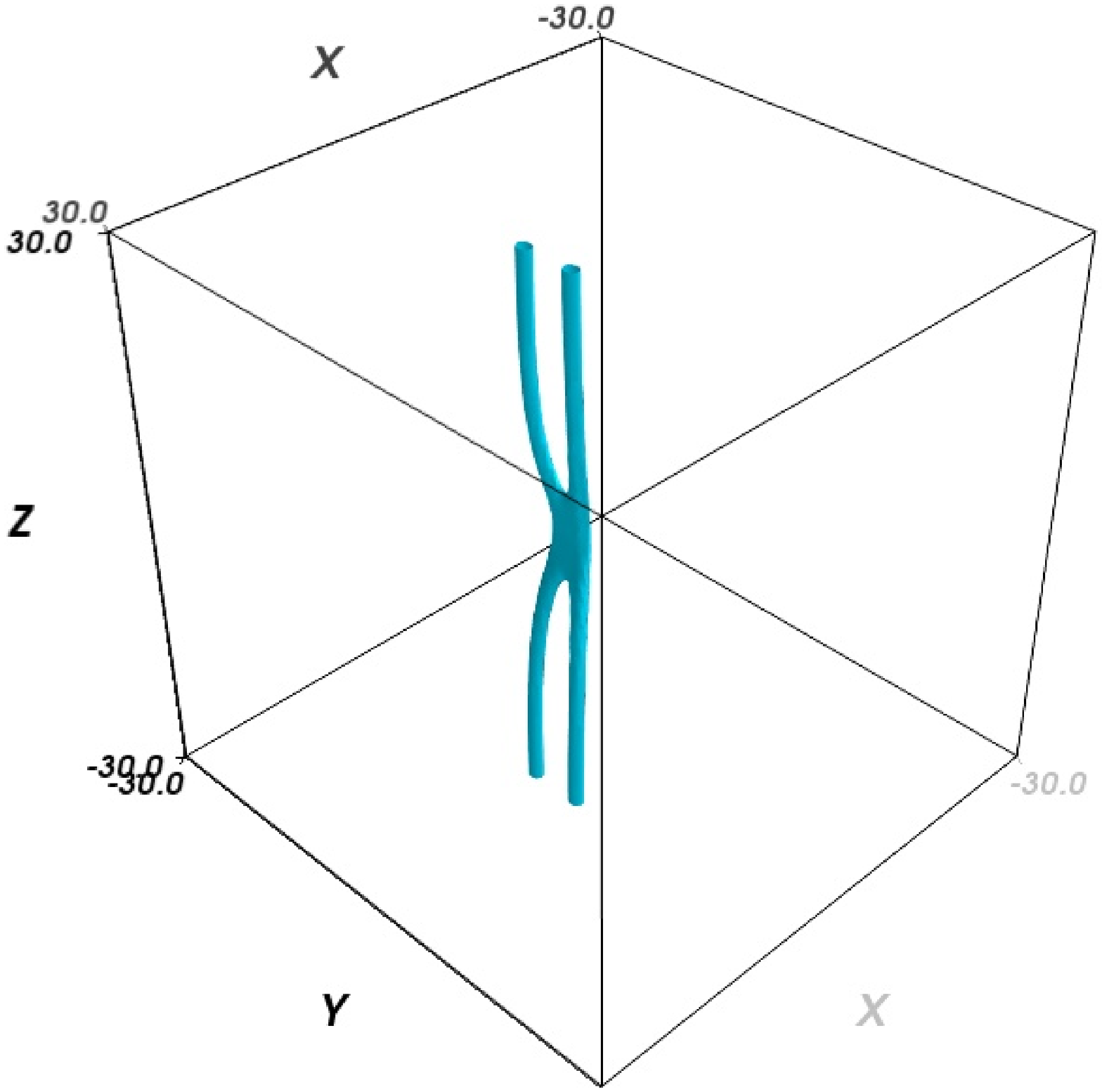}\\
\includegraphics[scale=.25]{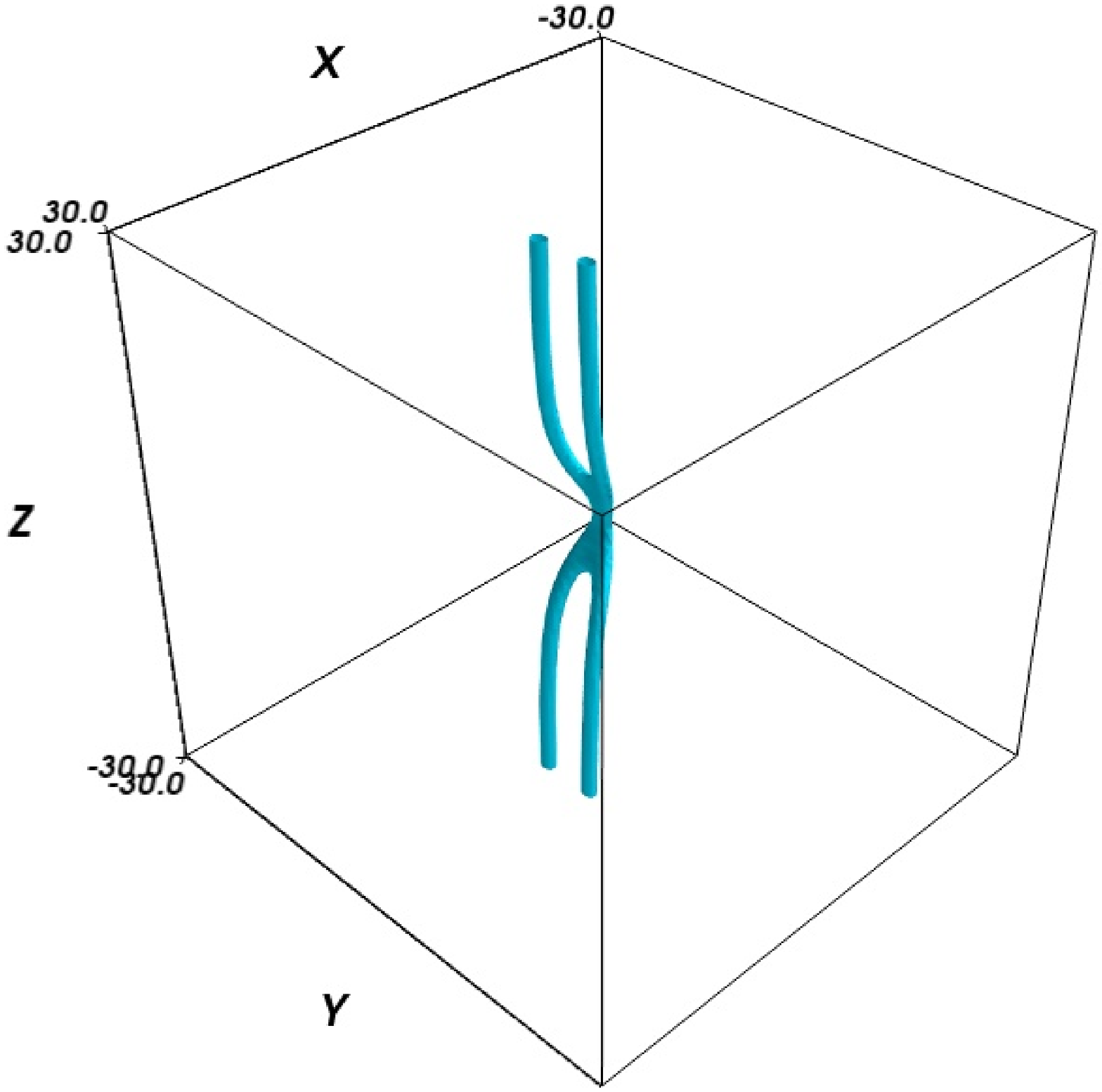} 
\includegraphics[scale=.25]{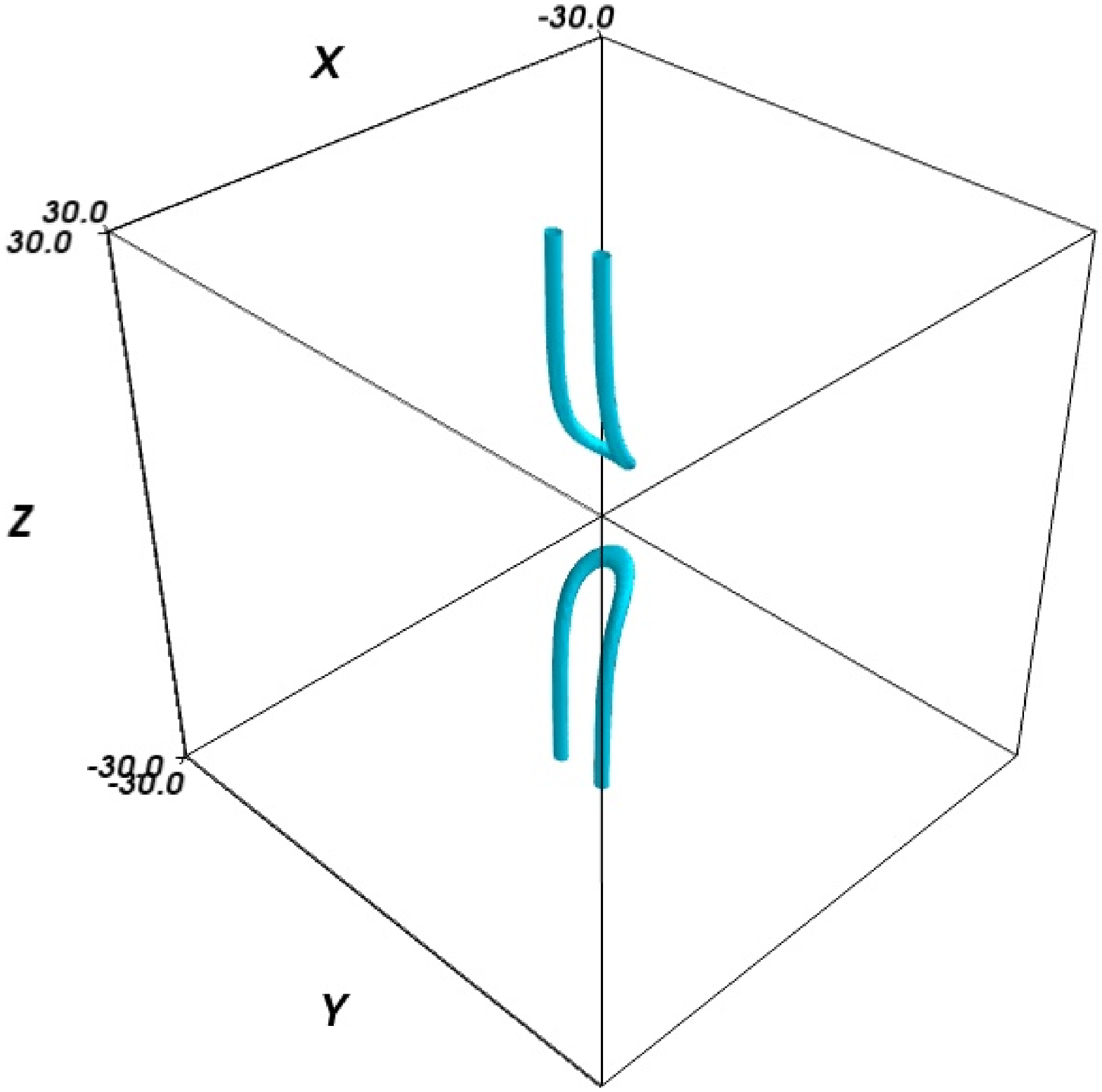}
\end{center}
\caption{Snapshots of the evolution of two anti-parallel vortices
(angle between vortices $\beta=\pi$), initially slightly
perturbed to enhance the Crow instability, at
$t=0$ (top left), $t=20$ (top right), $t=30$ (bottom left), $t=40$ (bottom
right).
Isosurfaces of $\rho = 0.2$ are plotted to visualise the vortex cores.}
\label{fig:3}
\end{figure}

\newpage

\begin{figure}[h!!!]
\begin{center}
\includegraphics[scale=.85]{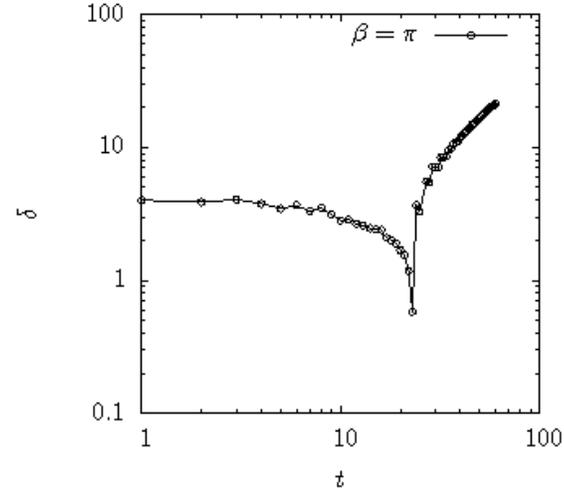}
\end{center}
\caption{Minimum distance $\delta$ between reconnecting
vortices as a function of time $t$ for the pair
of reconnecting anti-parallel vortices ($\beta=\pi$) shown in 
figure~(\ref{fig:3}).}
\label{fig:4}
\end{figure}

\newpage

\begin{figure}[h!!!]
\begin{center}
\includegraphics[scale=.25]{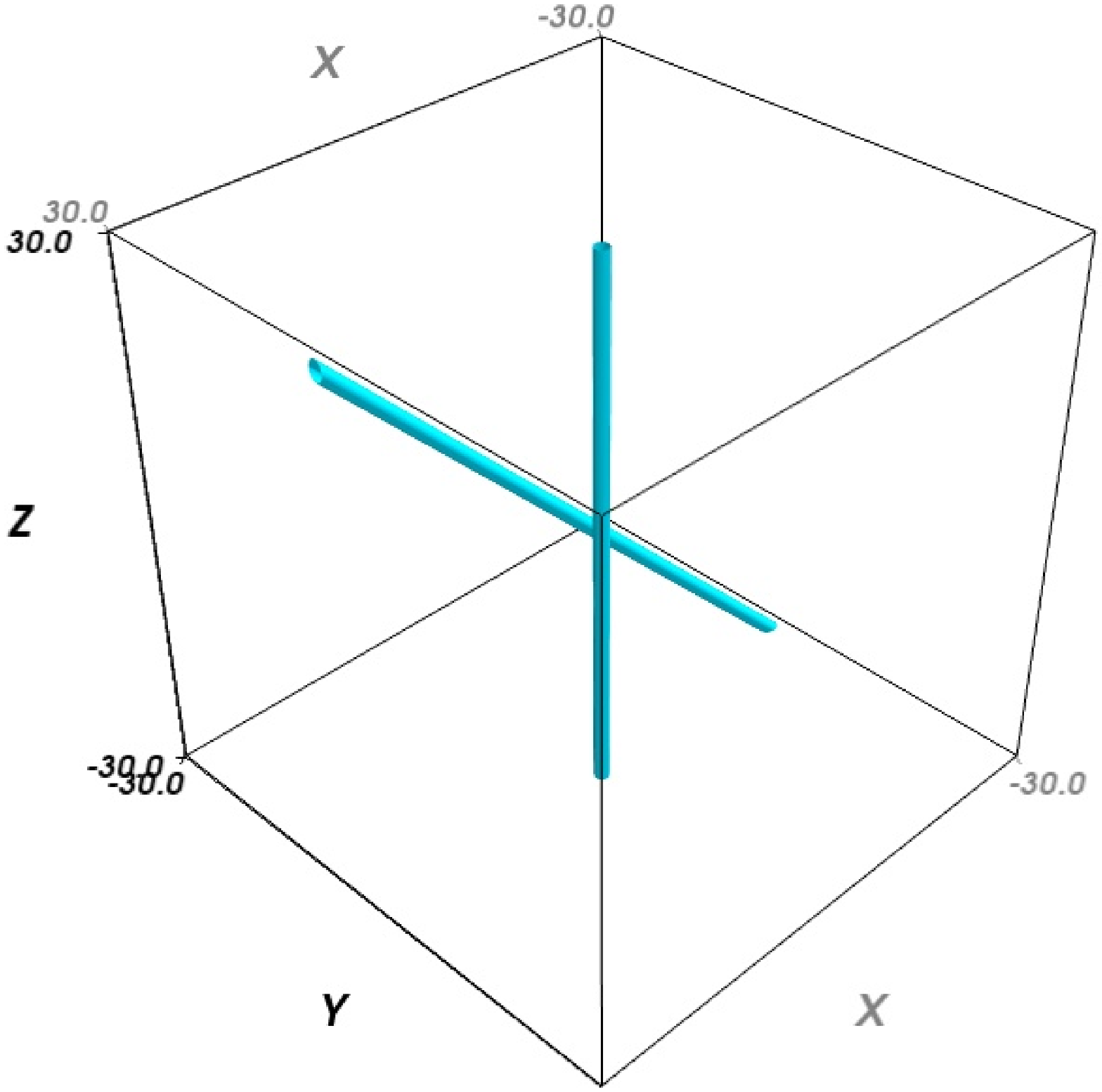} 
\includegraphics[scale=.25]{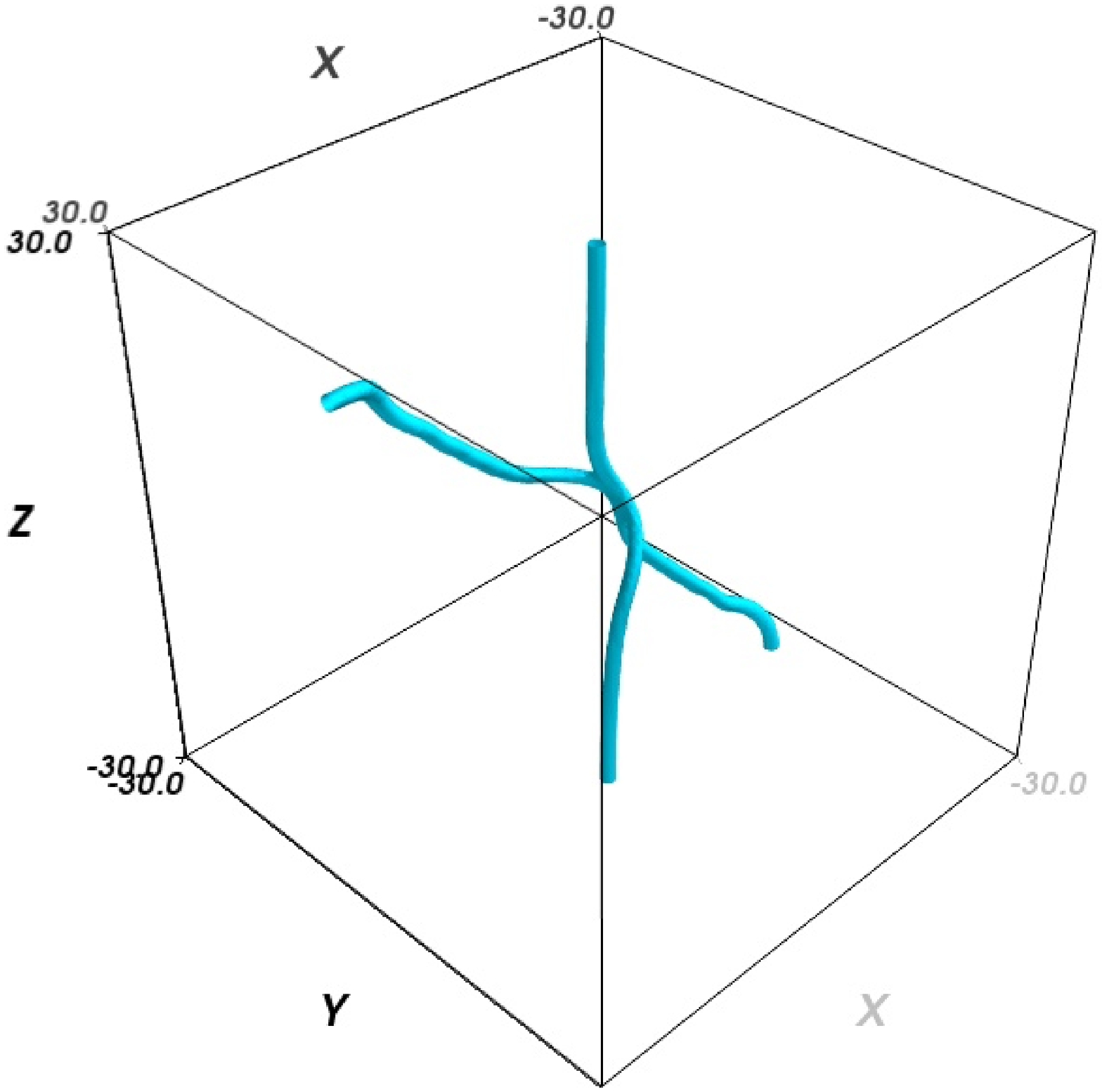}\\
\includegraphics[scale=.25]{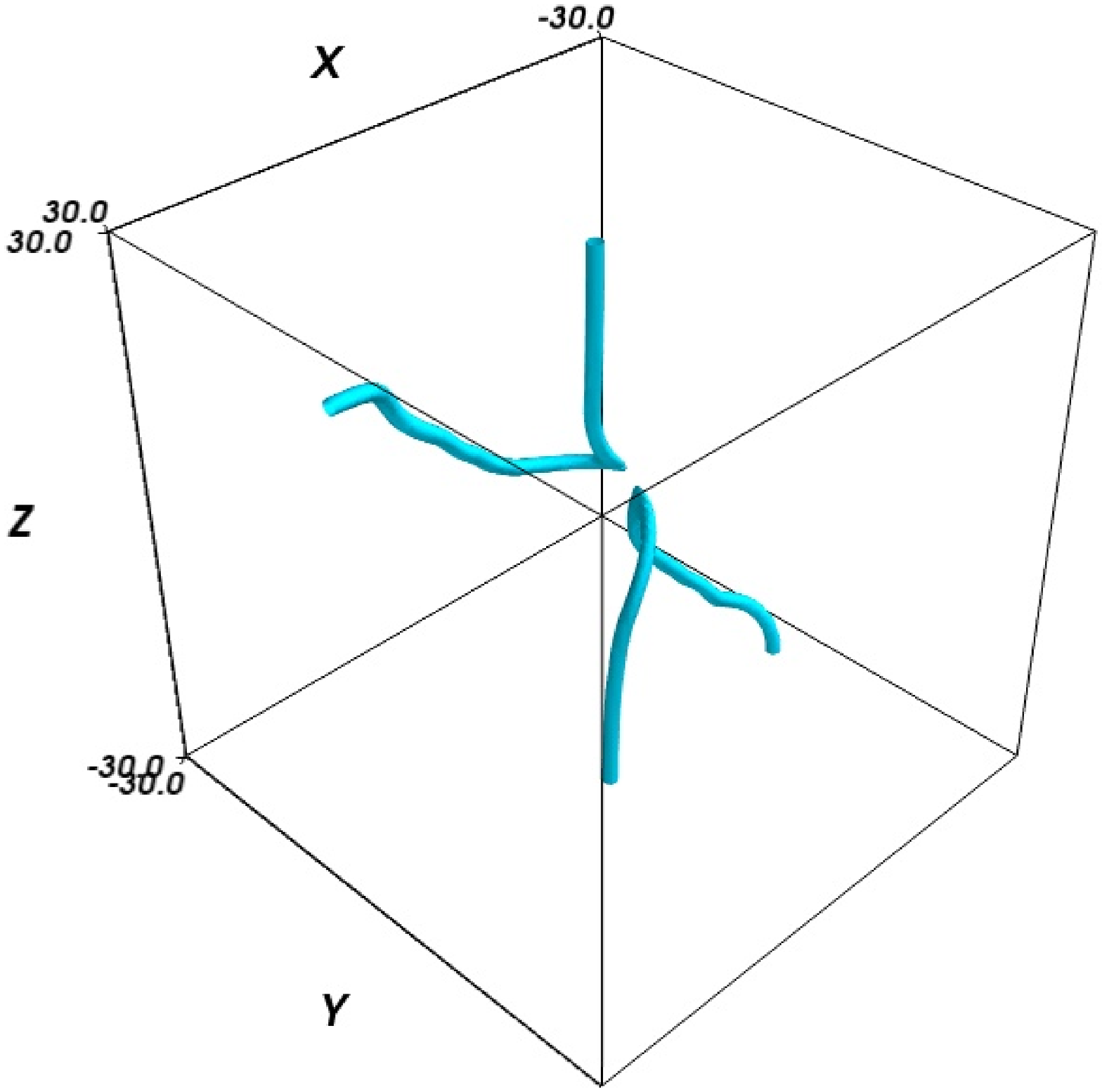} 
\includegraphics[scale=.25]{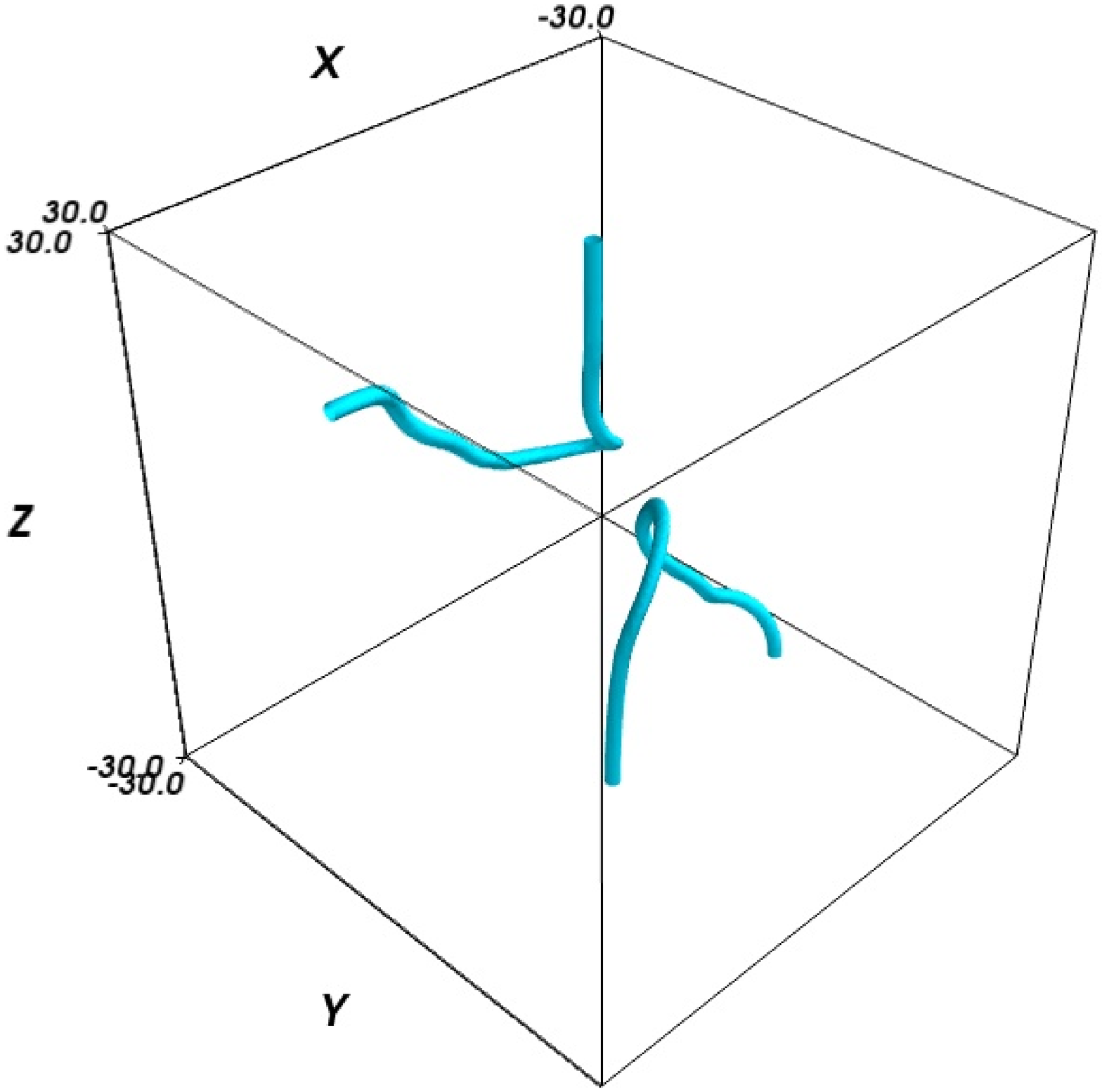}\\
\end{center}
\caption{Snapshots of the evolution of two initially straight vortices forming
an angle $\beta=3 \pi/4$ at
$t=0$ (top left), $t=15$ (top right), $t=20$ (bottom left), $t=25$ (bottom
right). 
Isosurfaces of $\rho=0.2$ are plotted to visualise the vortex cores.}
\label{fig:5}
\end{figure}

\newpage

\begin{figure}[h!!!]
\begin{center}
\includegraphics[scale=.25]{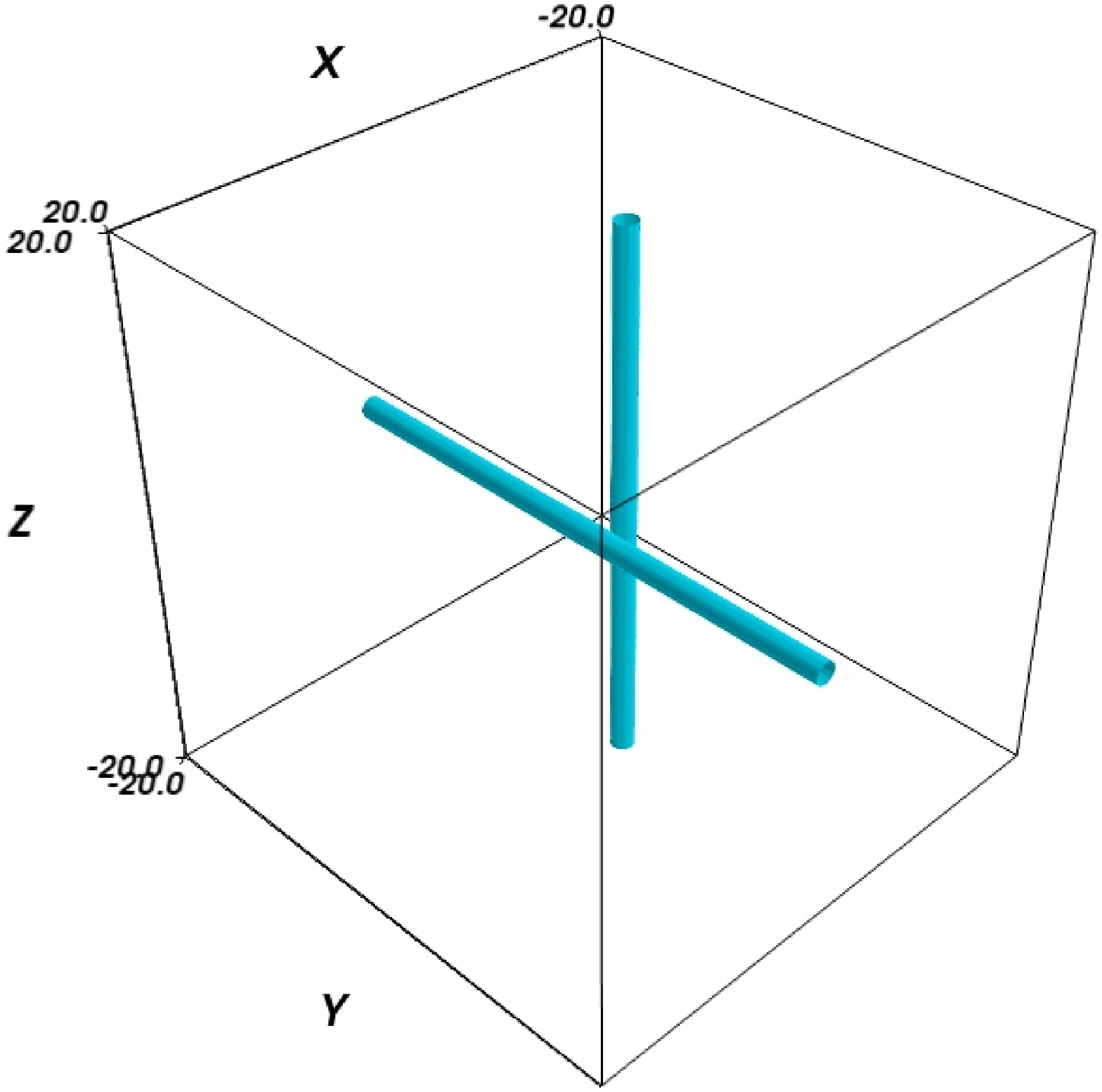} 
\includegraphics[scale=.25]{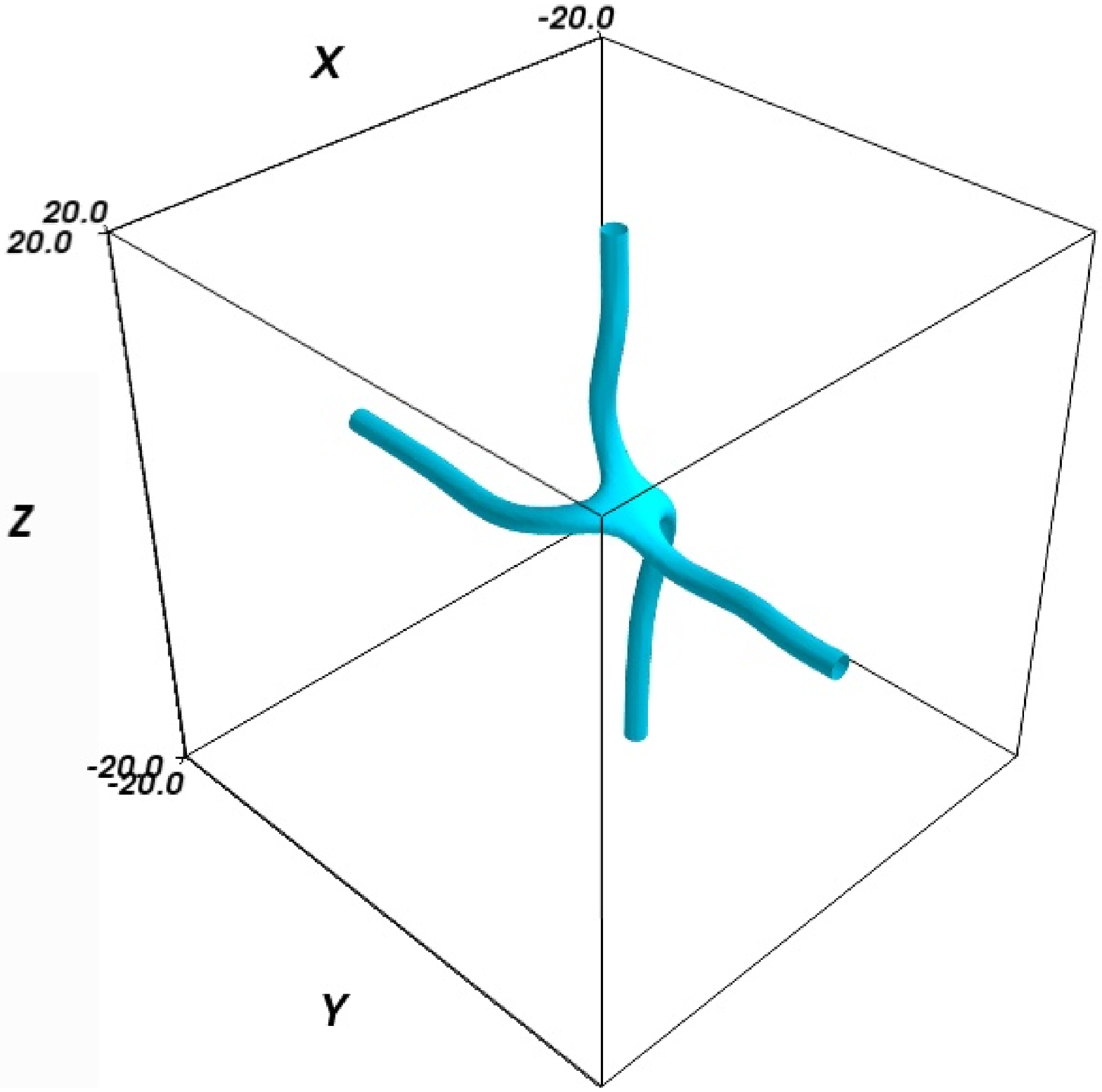}\\
\includegraphics[scale=.25]{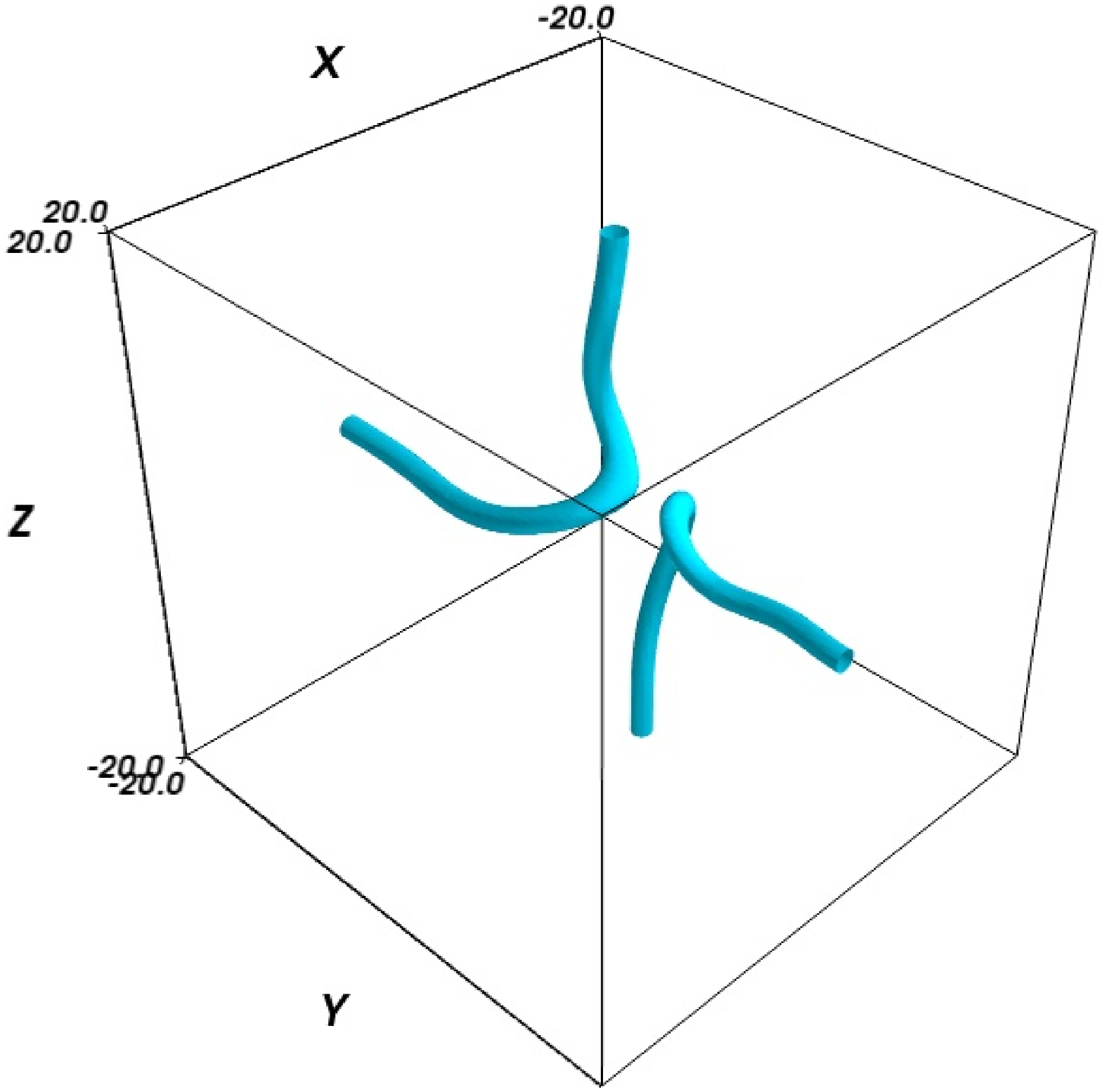} 
\includegraphics[scale=.25]{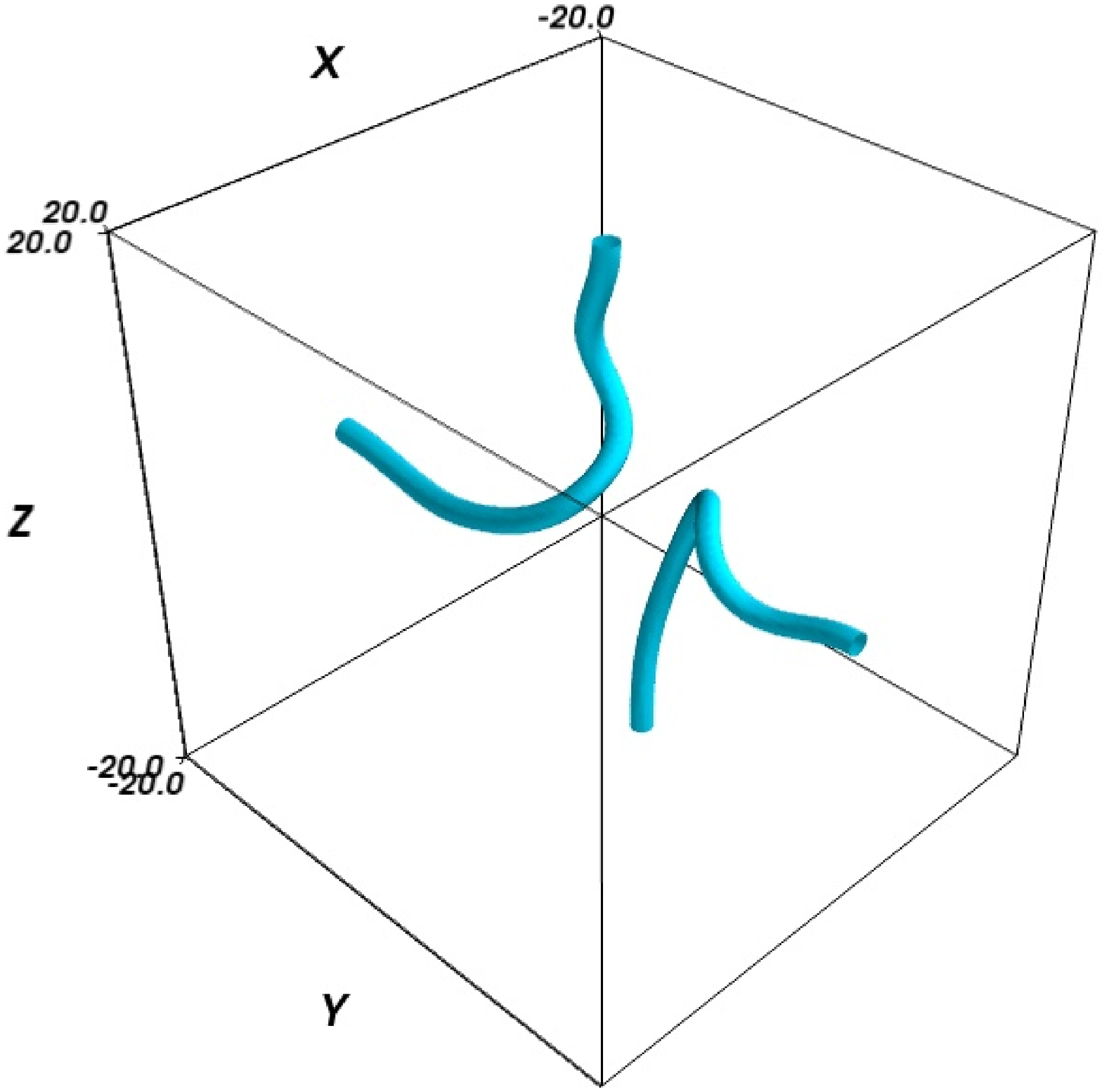}
\end{center}
\caption{Snapshots of the evolution of two perpendicular vortices
(angle between vortices $\beta=\pi/2$) at
$t=0$ (top left), $t=20$ (top right), $t=30$ (bottom left), $t=40$ (bottom
right).
Isosurfaces of $\rho=0.2$ are plotted to visualise the vortex cores.}
\label{fig:6}
\end{figure}

\newpage

\begin{figure}[h!!!]
\begin{center}
\includegraphics[scale=.91]{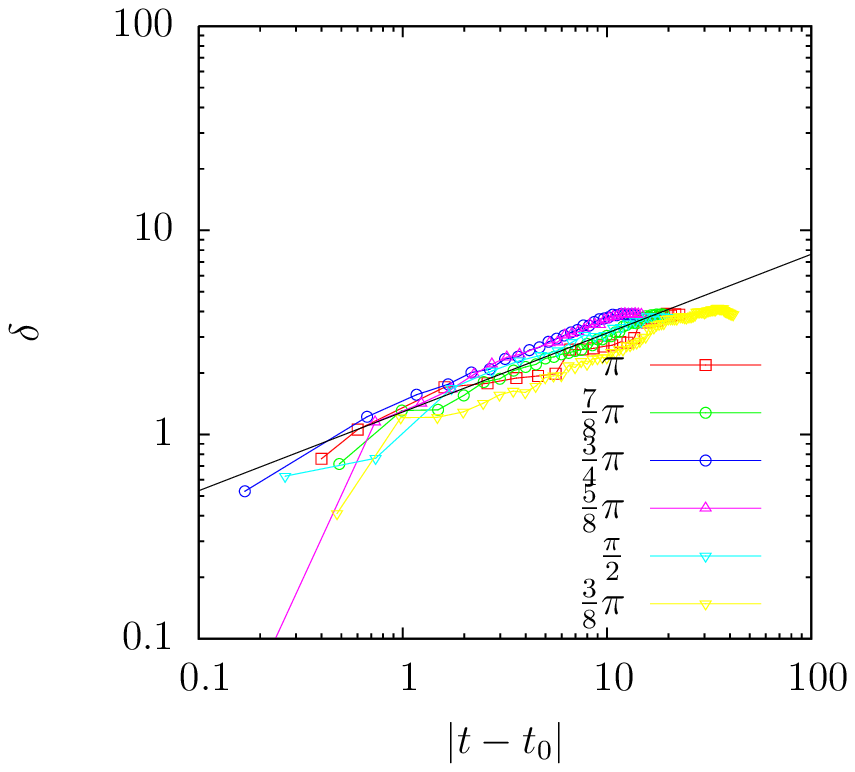} 
\includegraphics[scale=.91]{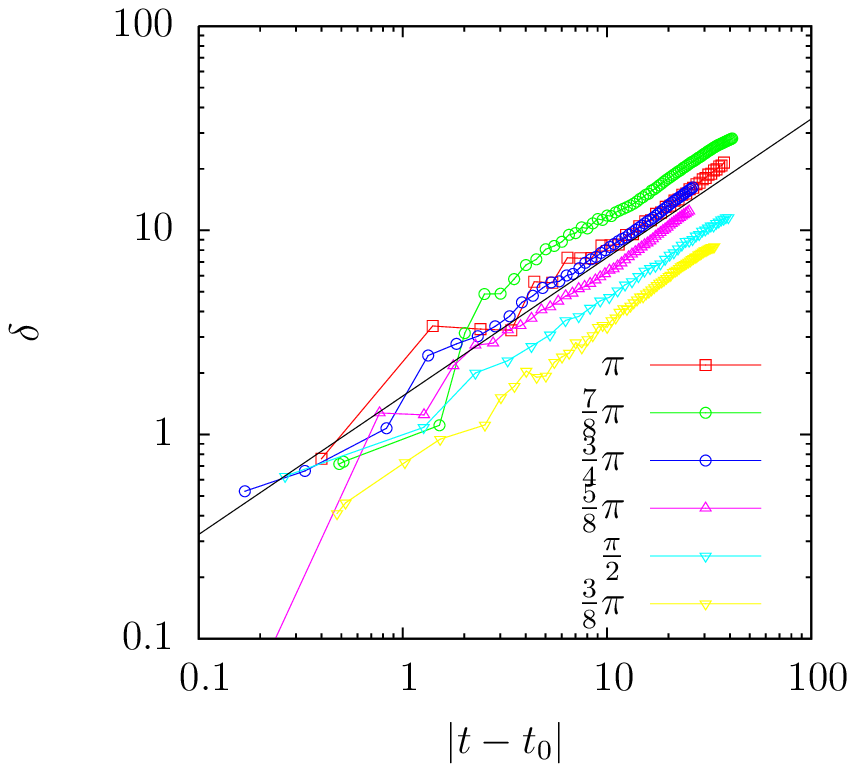}
\end{center}
\caption{Distance between vortices as a function of $\vert t-t_0 \vert$ 
before (top) and after (bottom) the reconnection for different values 
of the angle $\beta$ between initial
vortex lines ($\beta=\pi$ refers to anti-parallel vortices, and
$\beta=\pi/2$ to orthogonal vortices). The computed values are joined by
lines to guide the eye. The black solid lines are fits of the form
$\delta(t)=A \vert t-t_0 \vert^{\alpha}$; the fitting coefficients 
$A$ and $\alpha$ are reported in table I.
}
\label{fig:7}
\end{figure}

\newpage

\begin{figure}[h!!!]
\begin{center}
\includegraphics[scale=.25]{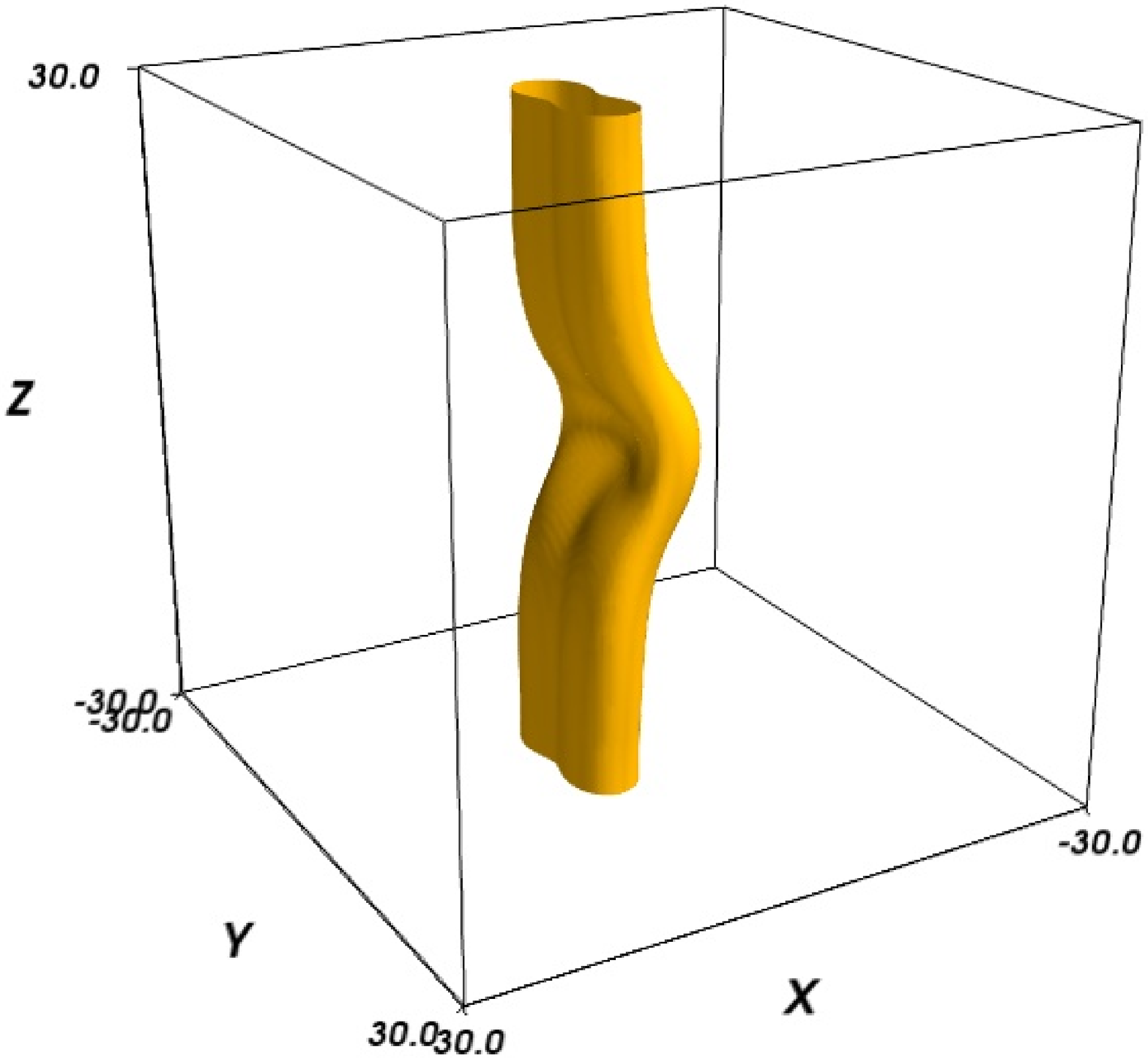} 
\includegraphics[scale=.25]{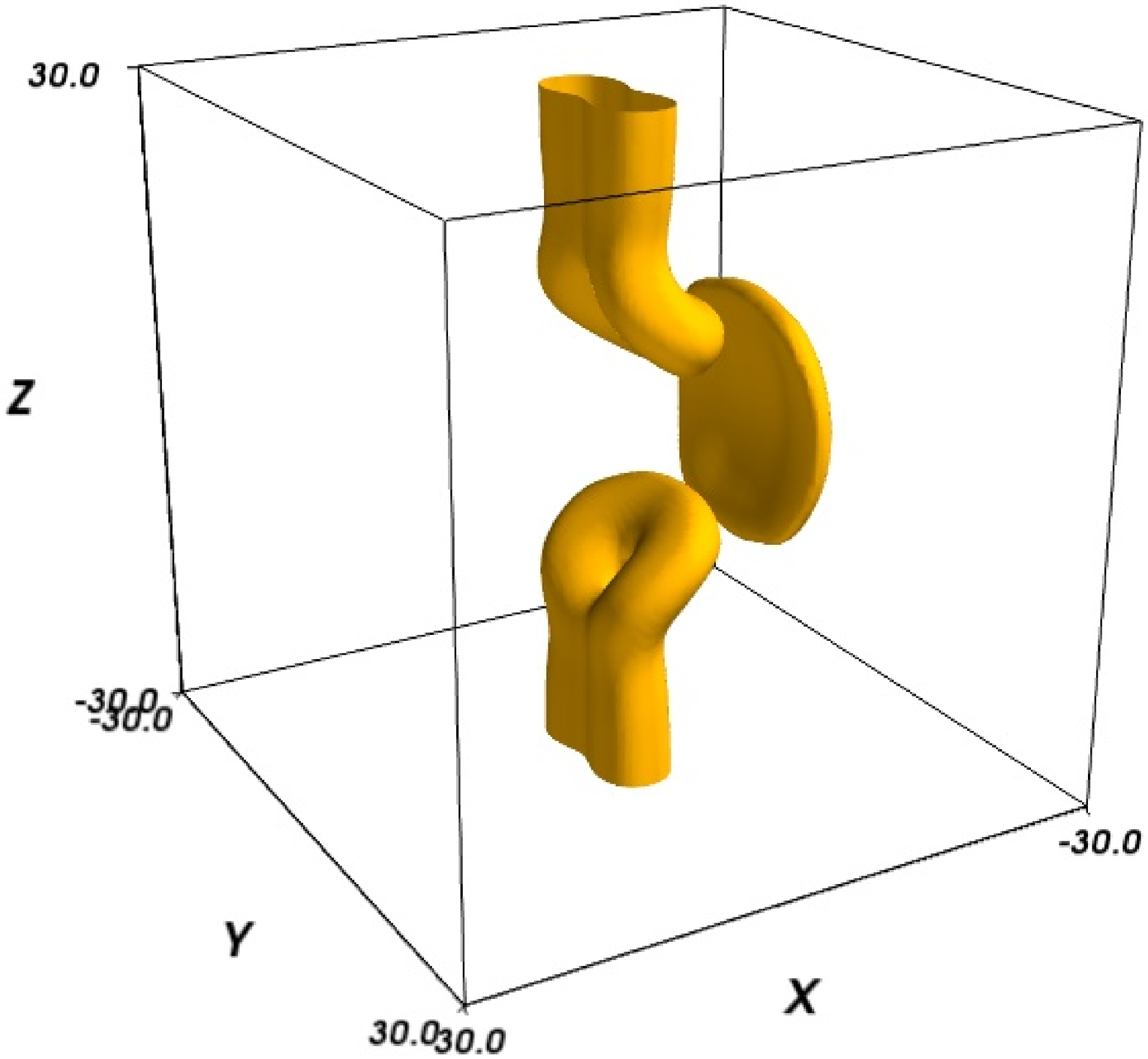}\\
\includegraphics[scale=.25]{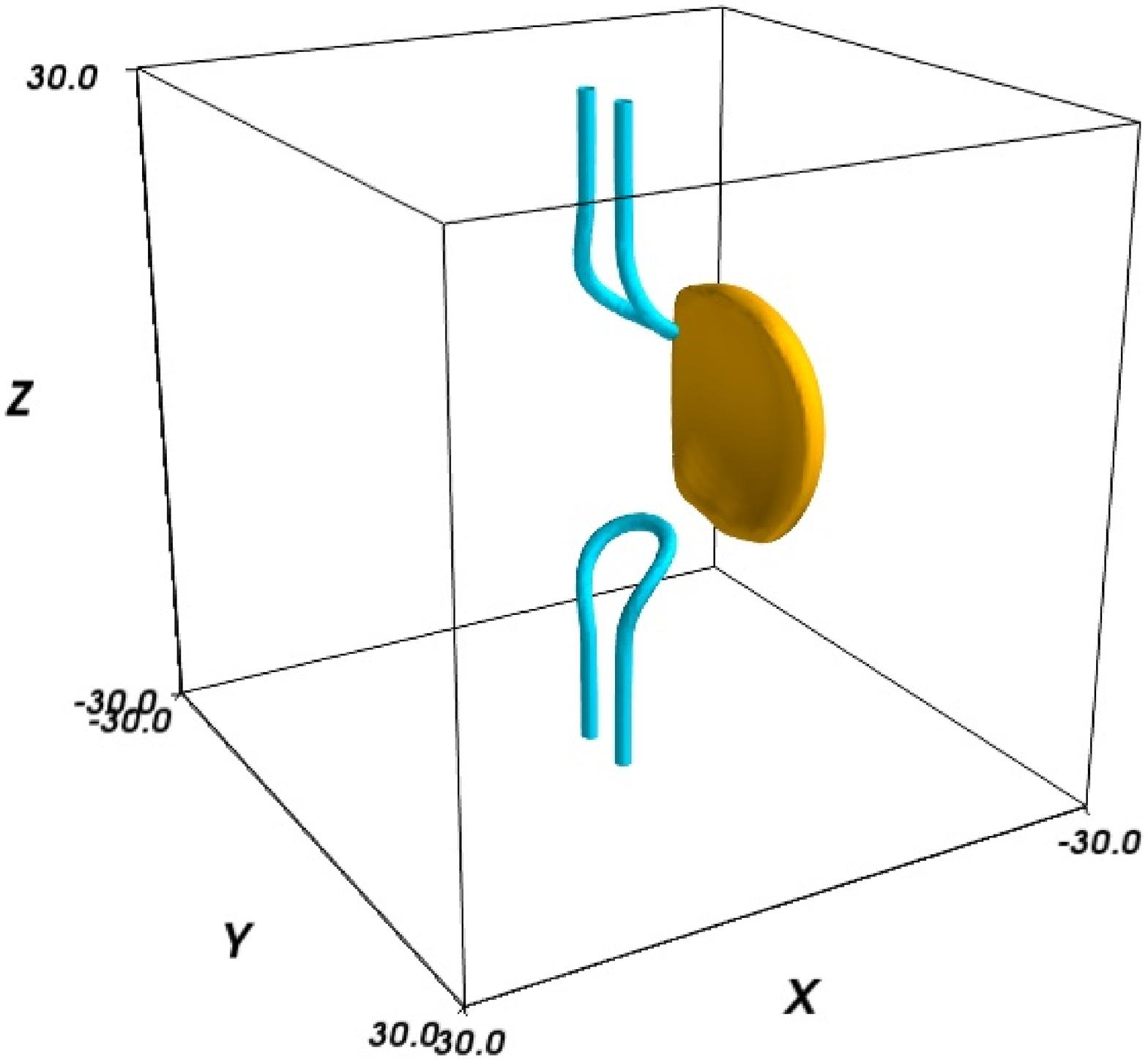}
\end{center}
\caption{Isosurfaces at $\rho=0.94$ before (left, $t=37$) and after (right and
bottom, $t=54$) reconnection for two anti-parallel vortices
(angle between vortices $\beta=\pi$). Note the mushroom-shaped 
rarefaction wave which moves away.}
\label{fig:8}
\end{figure}

\newpage

\begin{figure}[h!!!]
\begin{center}
\includegraphics[scale=.5]{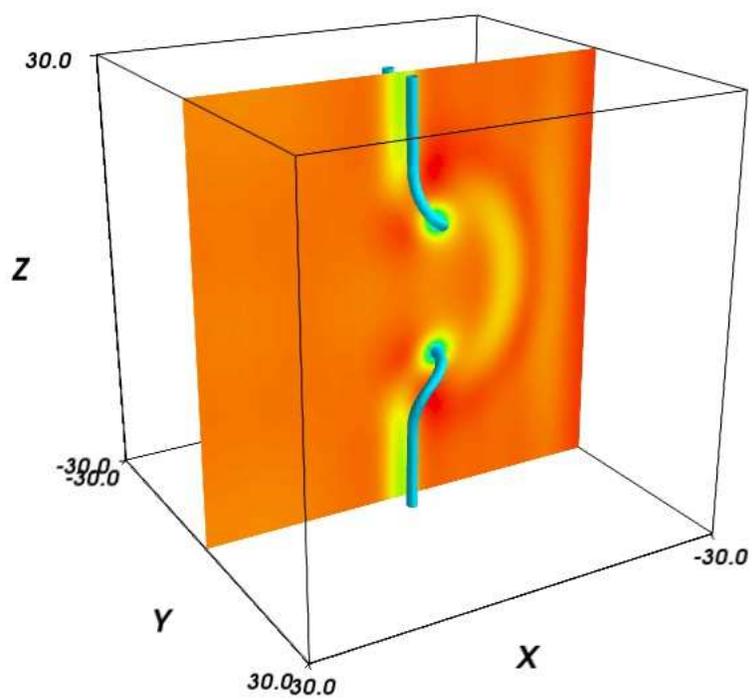} 
\end{center}
\caption{Isosurfaces at $\rho$ and $t=54$ as in figure~(\ref{fig:8}) 
(bottom, anti-parallel vortices) to visualise the vortex core with
superimposed the profile of $\rho$ on the $y=0$ plane. Note the
rarefaction wave.
}
\label{fig:9}
\end{figure}

\newpage

\begin{figure}[h!!!]
\begin{center}
\includegraphics[scale=.25]{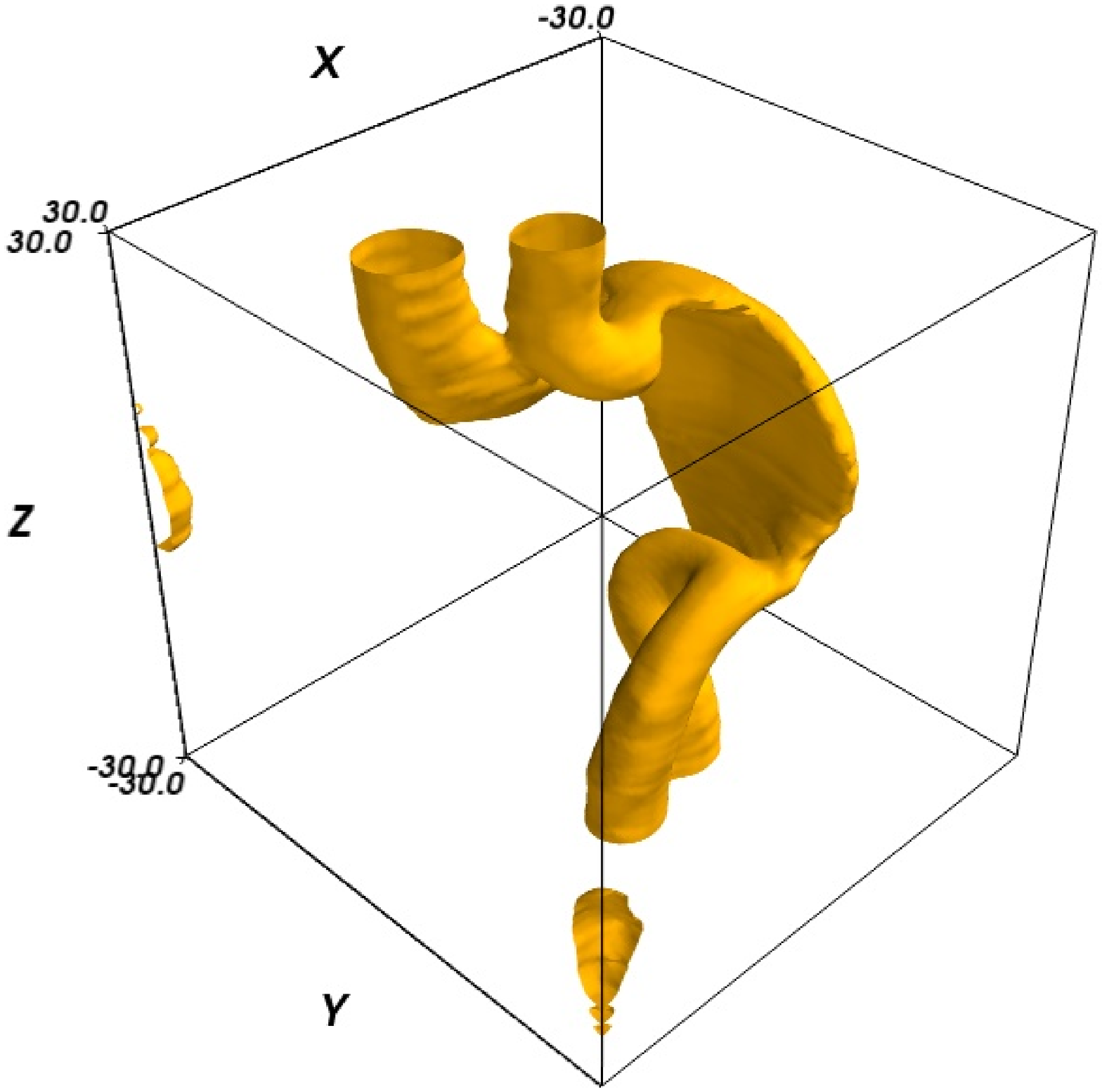} 
\includegraphics[scale=.25]{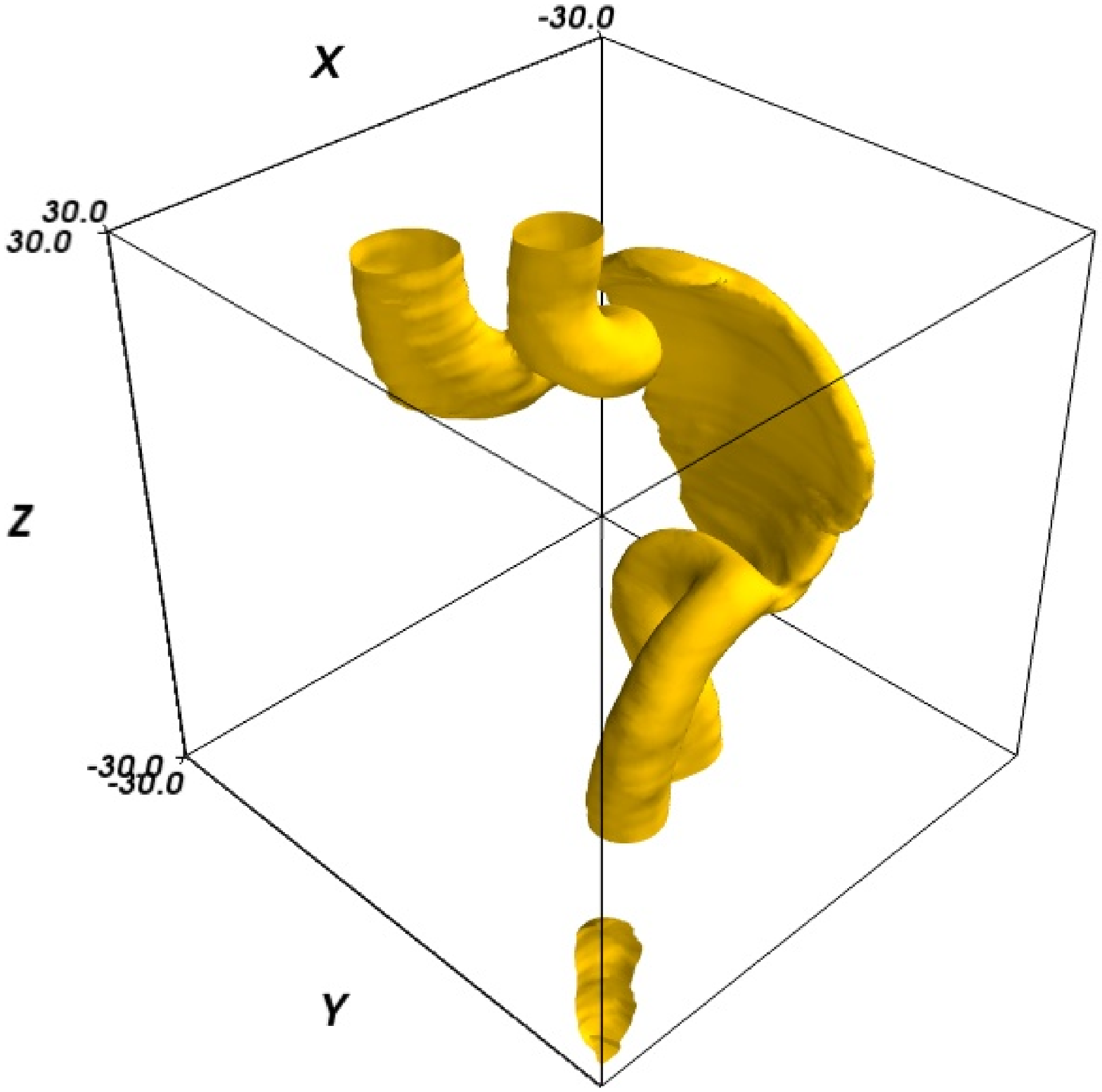}\\
\includegraphics[scale=.25]{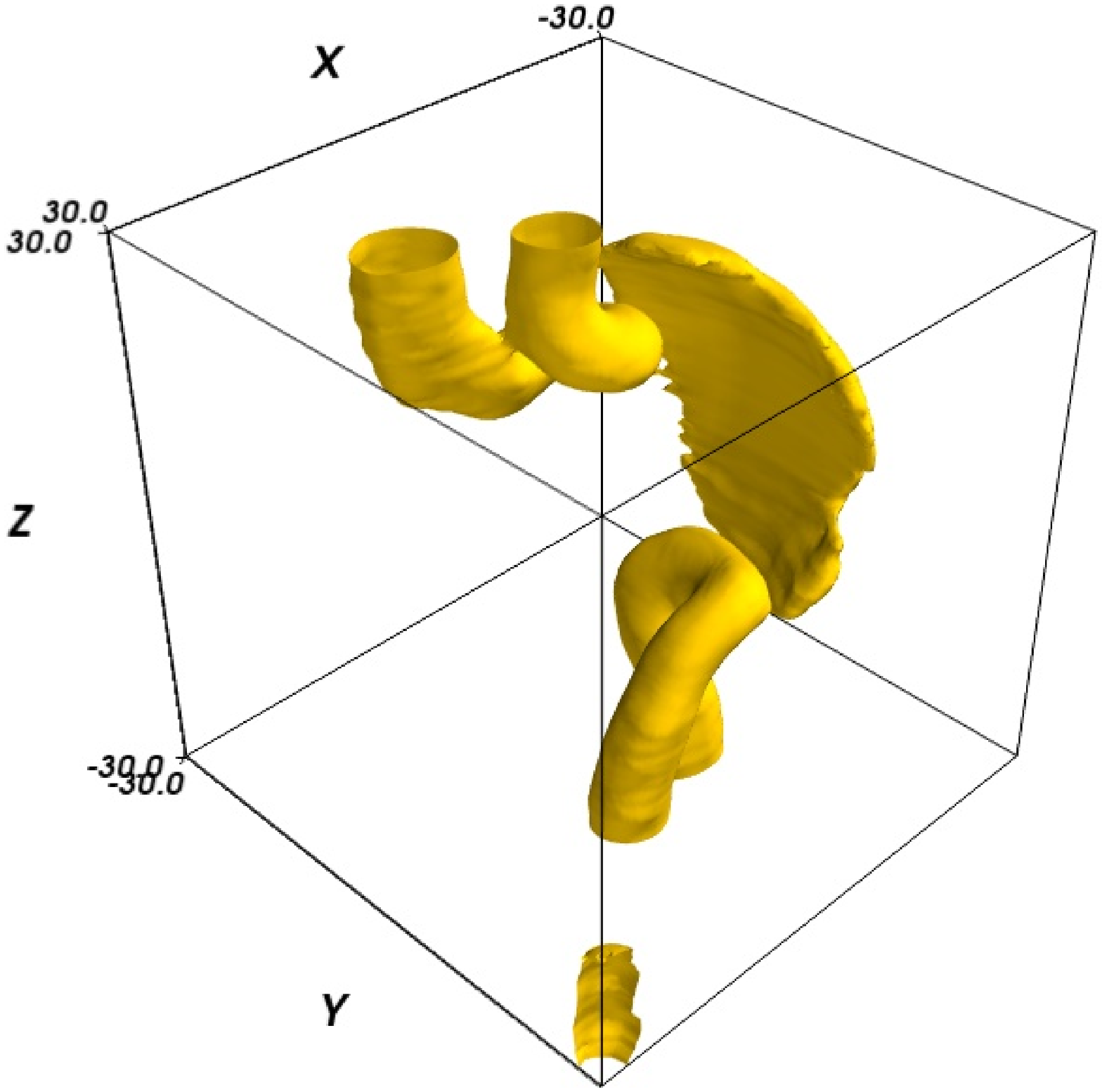} 
\includegraphics[scale=.25]{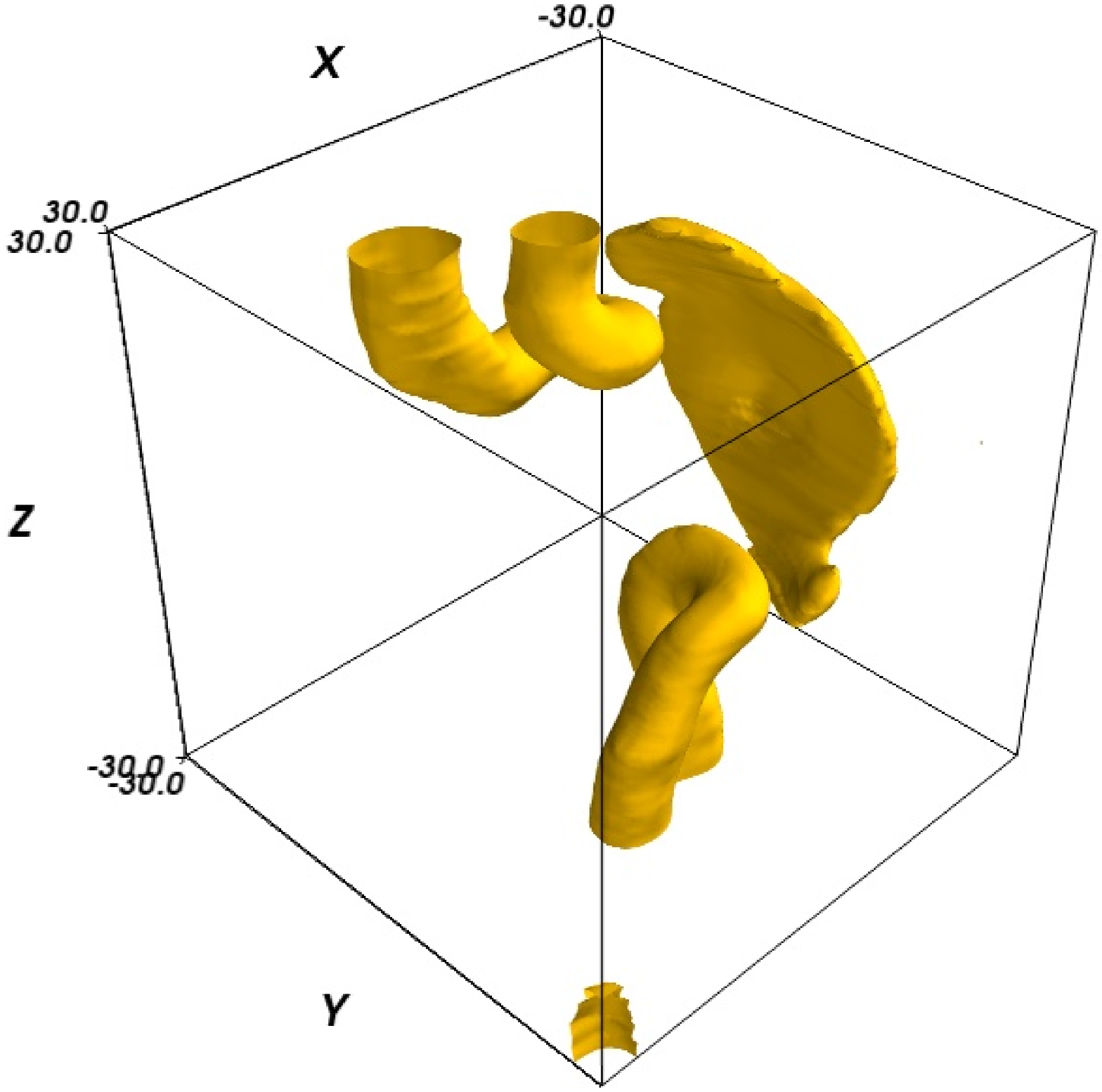}\\
\end{center}
\caption{Snapshots of the evolution of the mushroom-shaped rarefaction 
wave ejected after reconnection, $\beta= 7 \pi / 8$, at
$t=53$ (top left), $t=55$ (top right), $t=57$ (bottom left), $t=59$ (bottom
right); 
isosurfaces of $\rho=0.94$ are plotted to visualize the vortex cores
and the wave as in figure~(\ref{fig:8}).}
\label{fig:10}
\end{figure}

\newpage

\begin{figure}[h!!!]
\begin{center}
\includegraphics[scale=.25]{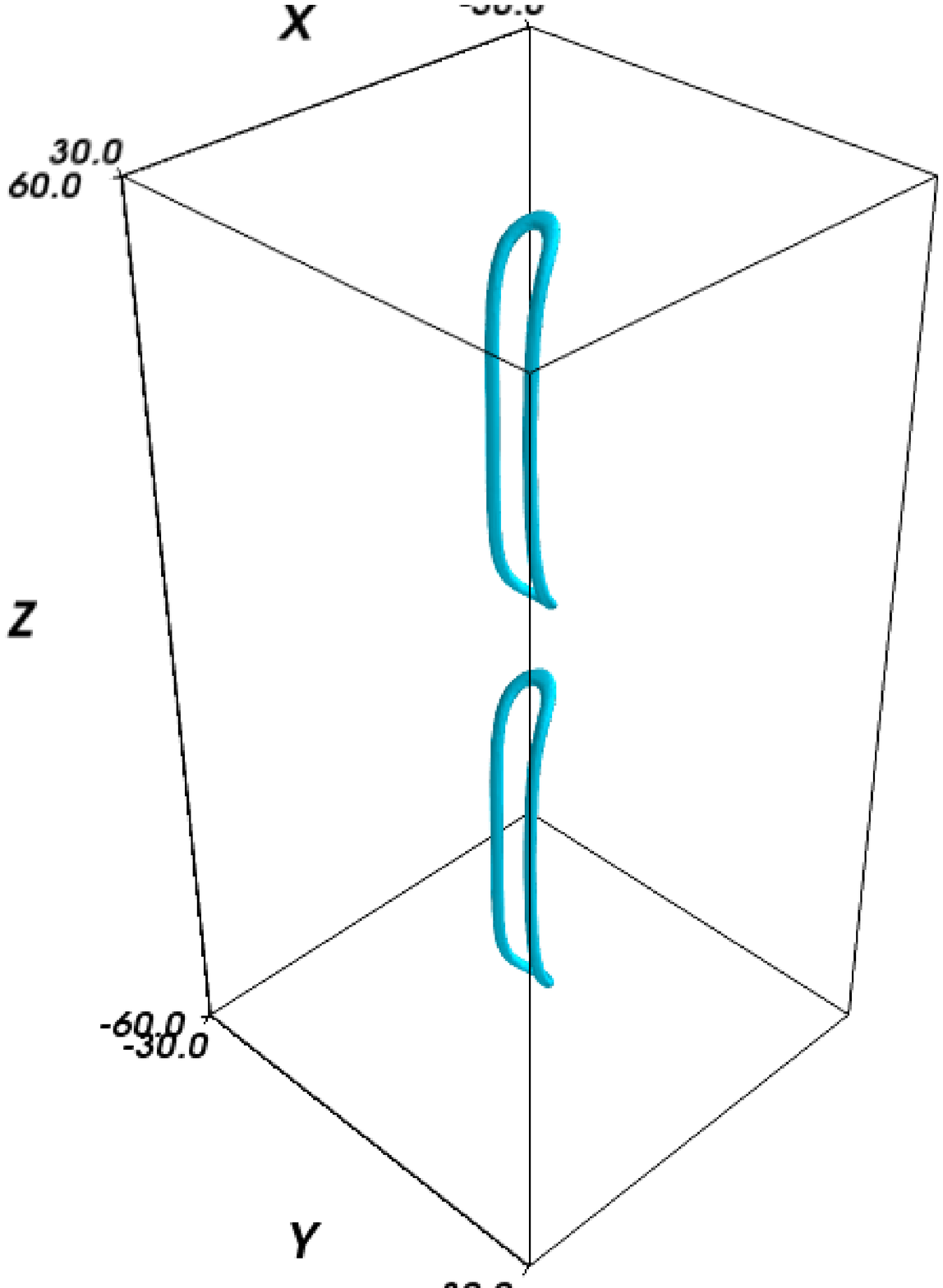} 
\includegraphics[scale=.25]{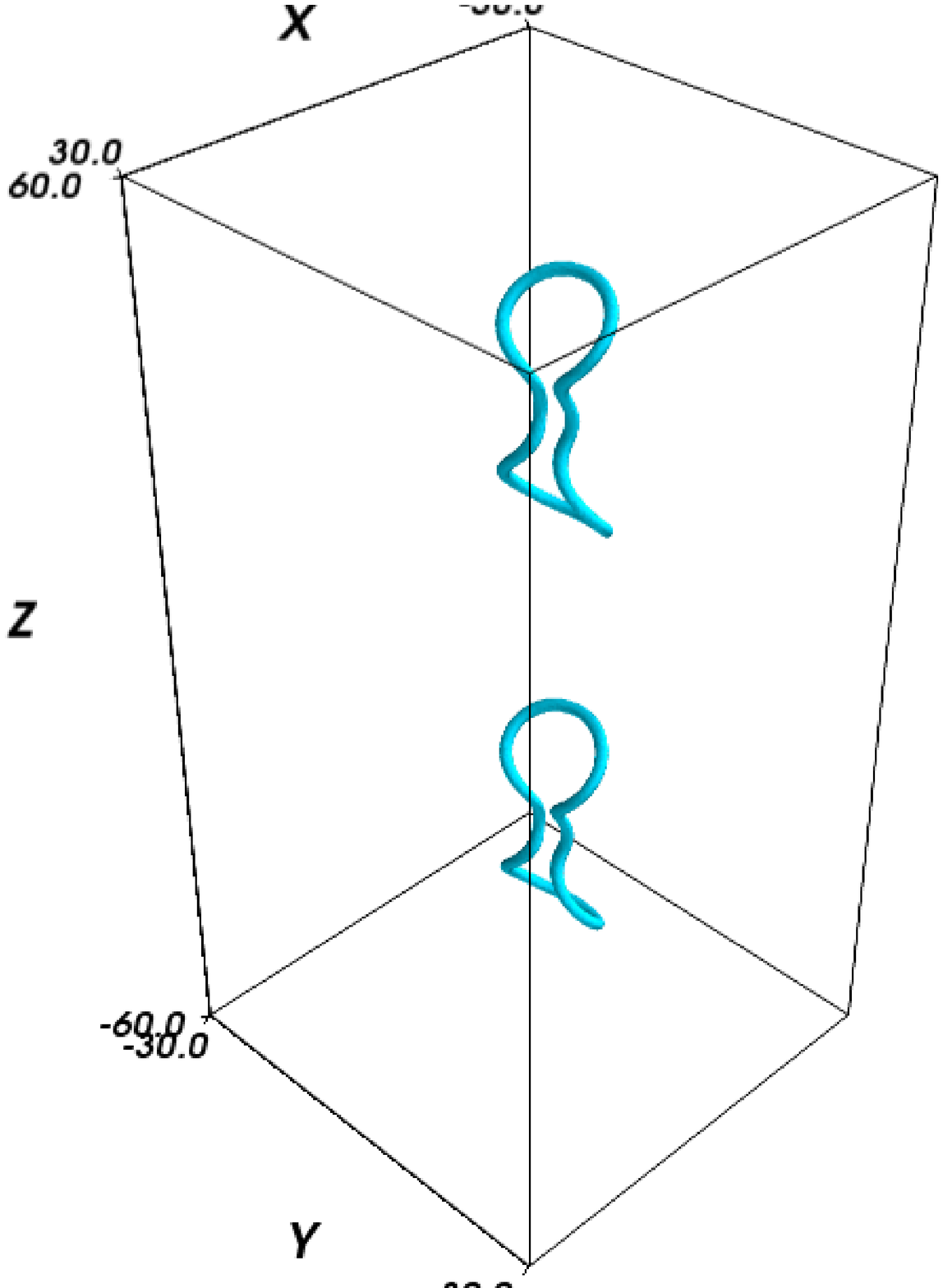}\\
\includegraphics[scale=.25]{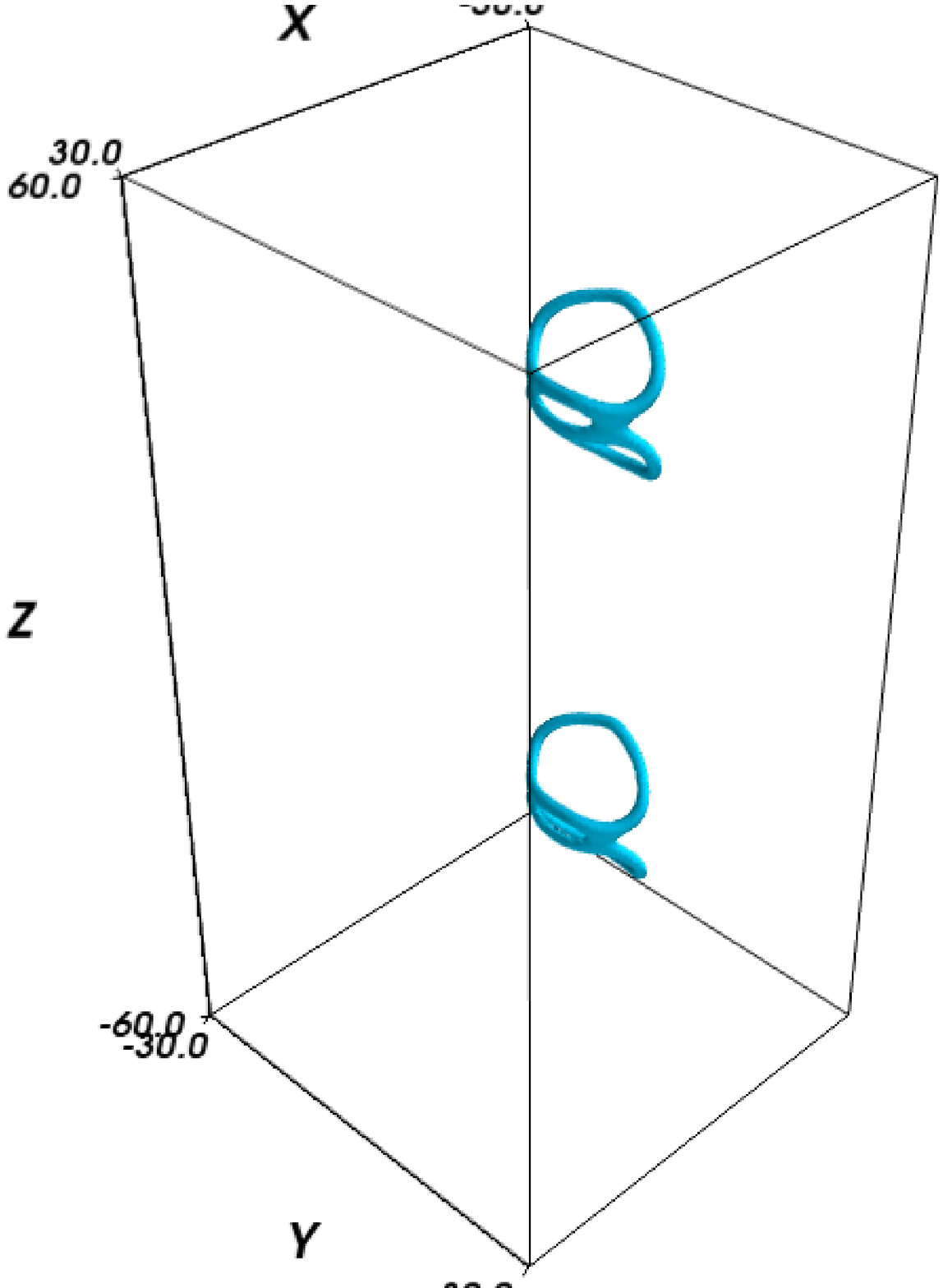} 
\includegraphics[scale=.25]{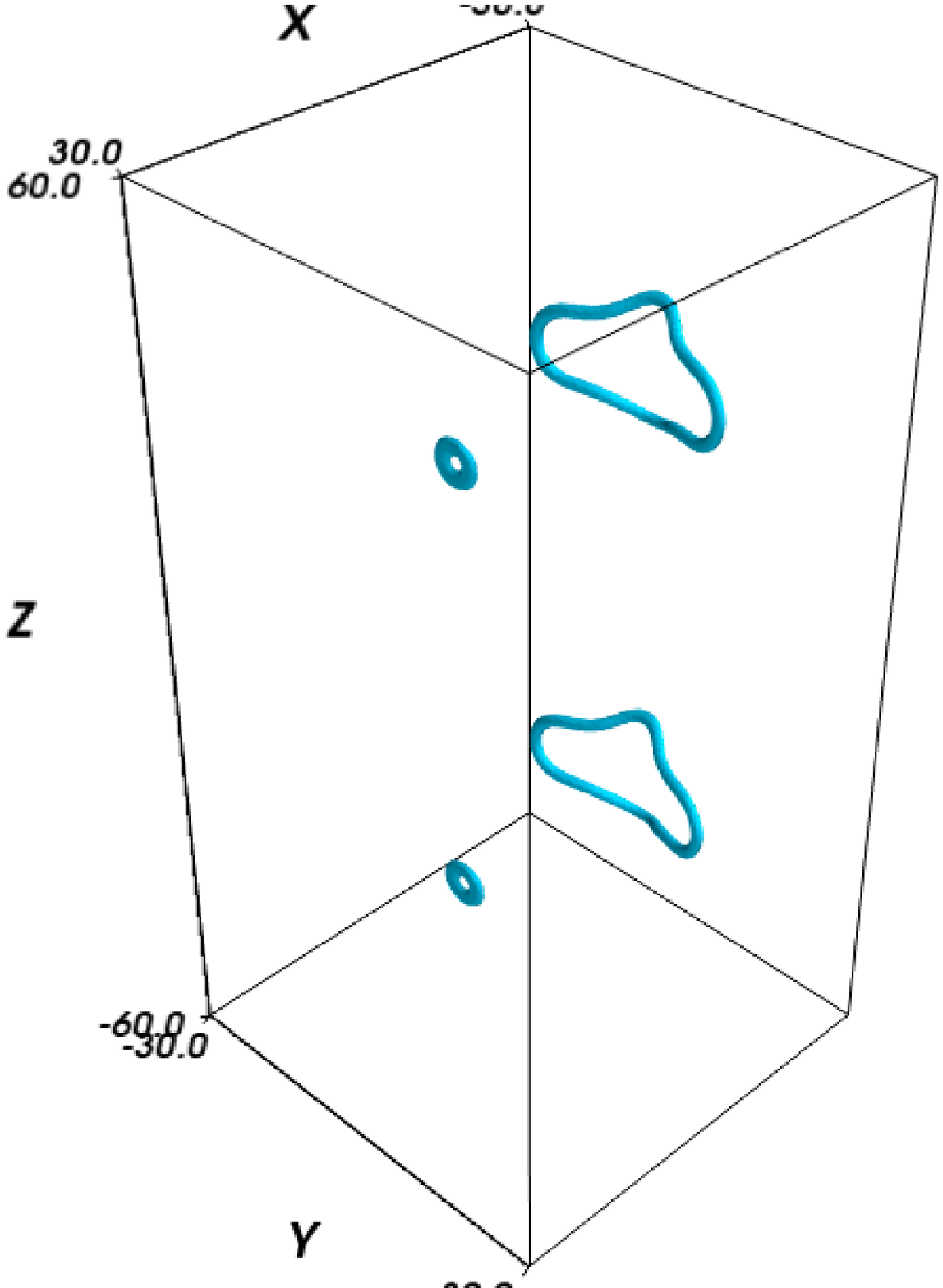}\\
\end{center}
\caption{
Snapshots of the evolution of two antiparallel vortices
(angle $\beta=\pi$) initially slightly perturbed to enhance the 
Crow instability, at $t=40$ (top left), $t=80$ (top right),
$t=120$ (bottom left) and $t=160$ (bottom right). Isosurface of
$\rho=0.2$ are plotted to visualize the vortex cores. The initial
condition is the same as in figure~\ref{fig:3} but the
computational box was extended to $-60 \leq z \leq 60$ to visualize
the formation of vortex rings. 
}
\label{fig:11}
\end{figure}

\newpage

\begin{figure}[h!!!]
\begin{center}
\includegraphics[scale=.39]{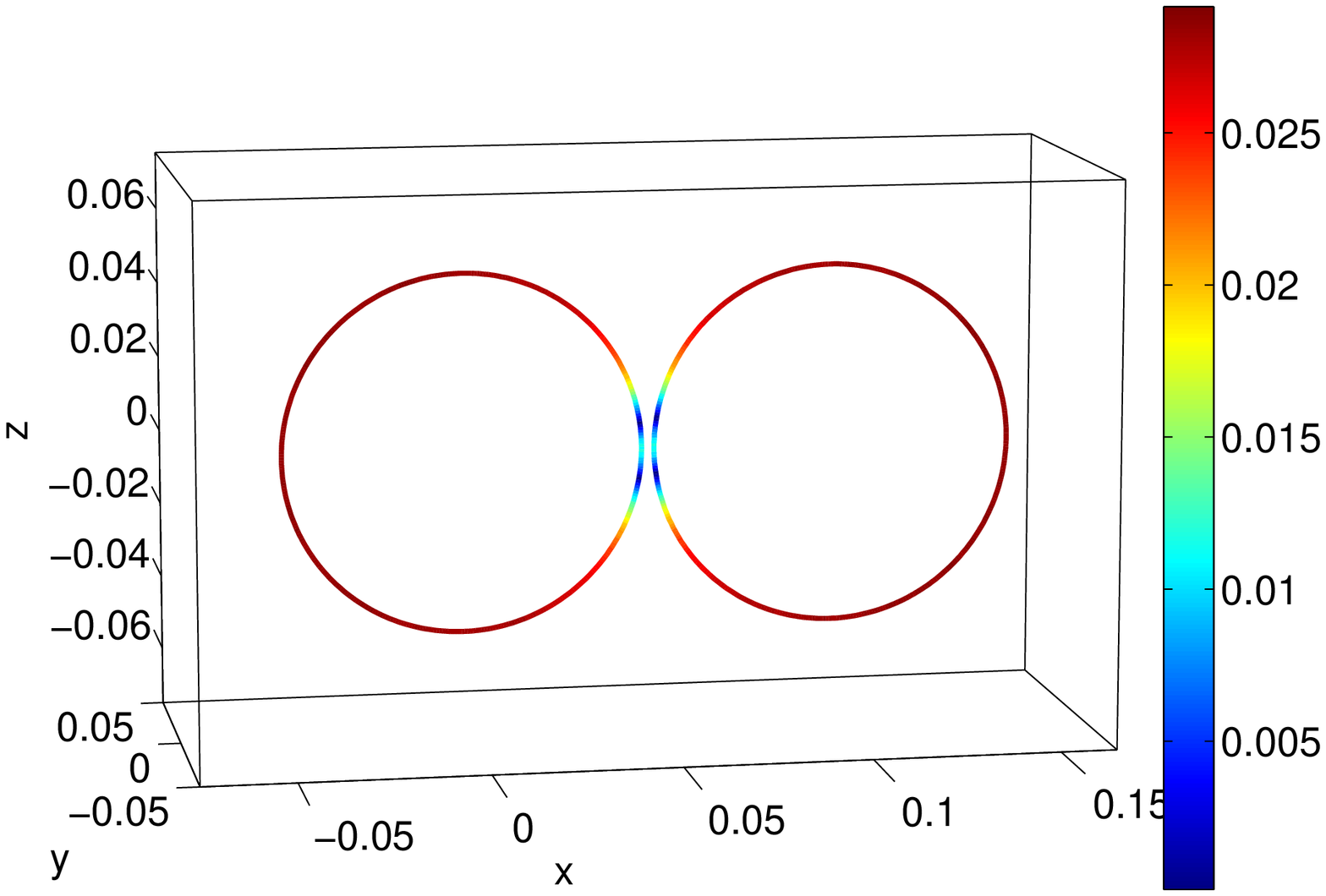}\\
\includegraphics[scale=.39]{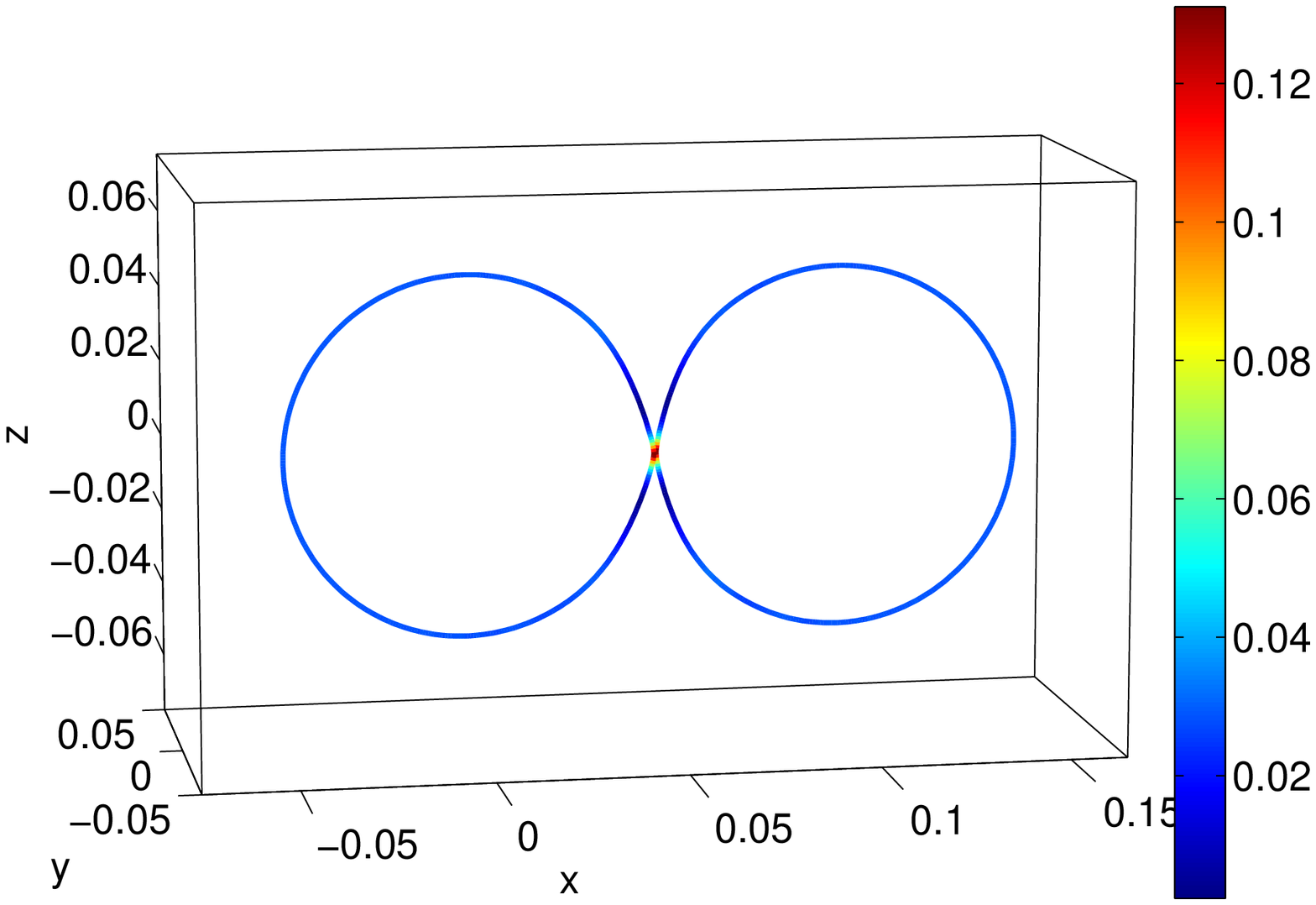}\\
\includegraphics[scale=.39]{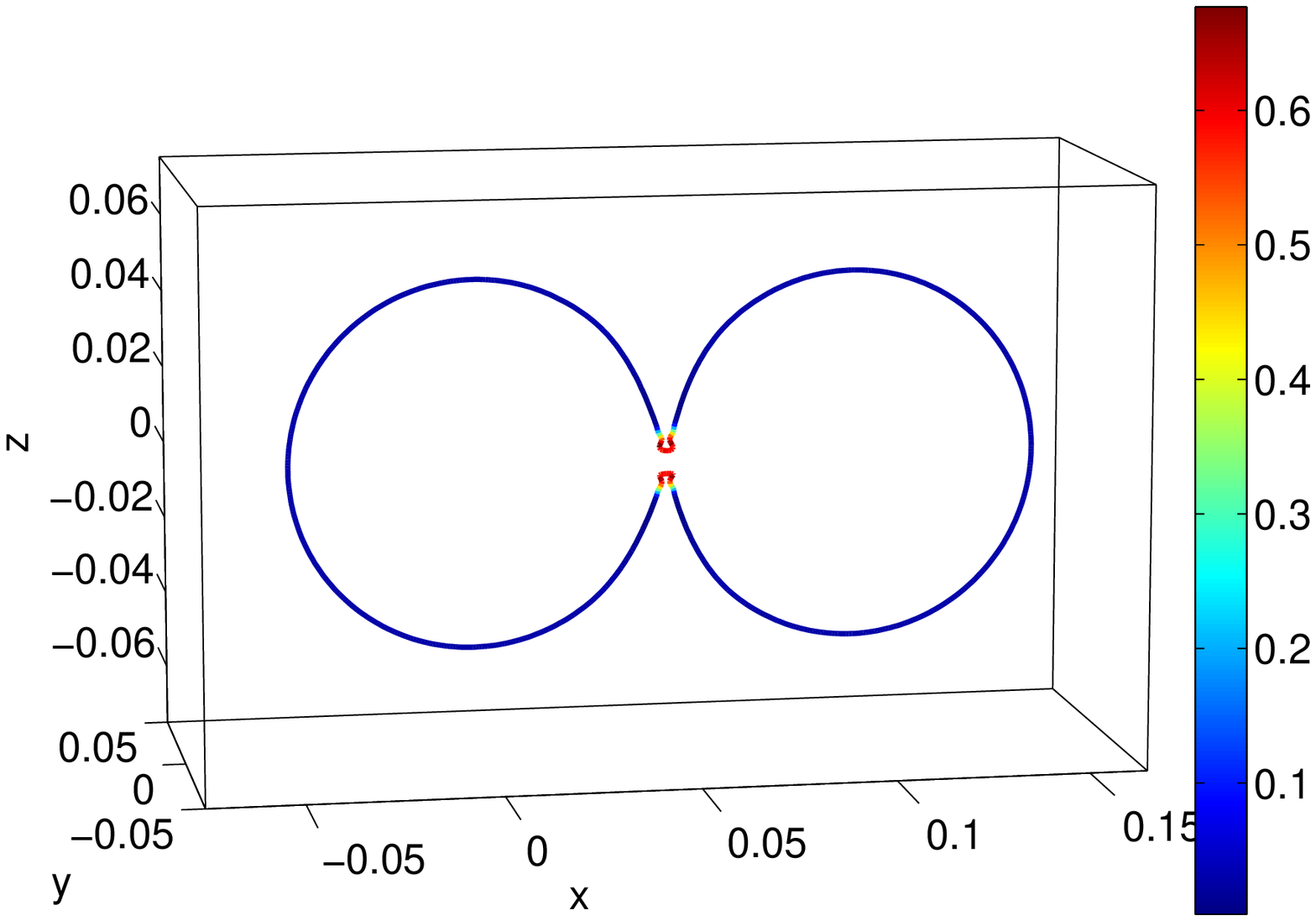} 
\end{center}
\caption{Reconnection of two vortex rings initially set parallel
to each other (as in the work of de Waele \& Aarts
\cite{WA1994}), computed with the Biot-Savart law:
Top: at time $t=0$;  middle: $t_0-t=0.001~\rm s$; 
bottom: $t-t_0=0.005~\rm s$. The vortex lines
are colour-coded to indicate the magnitude of the velocity (in $\rm cm/s$,
see legend on each figure). The box is for visualization only.
}
\label{fig:12}
\end{figure}

\newpage

\begin{figure}[h!!!]
\begin{center}
\includegraphics[scale=.39]{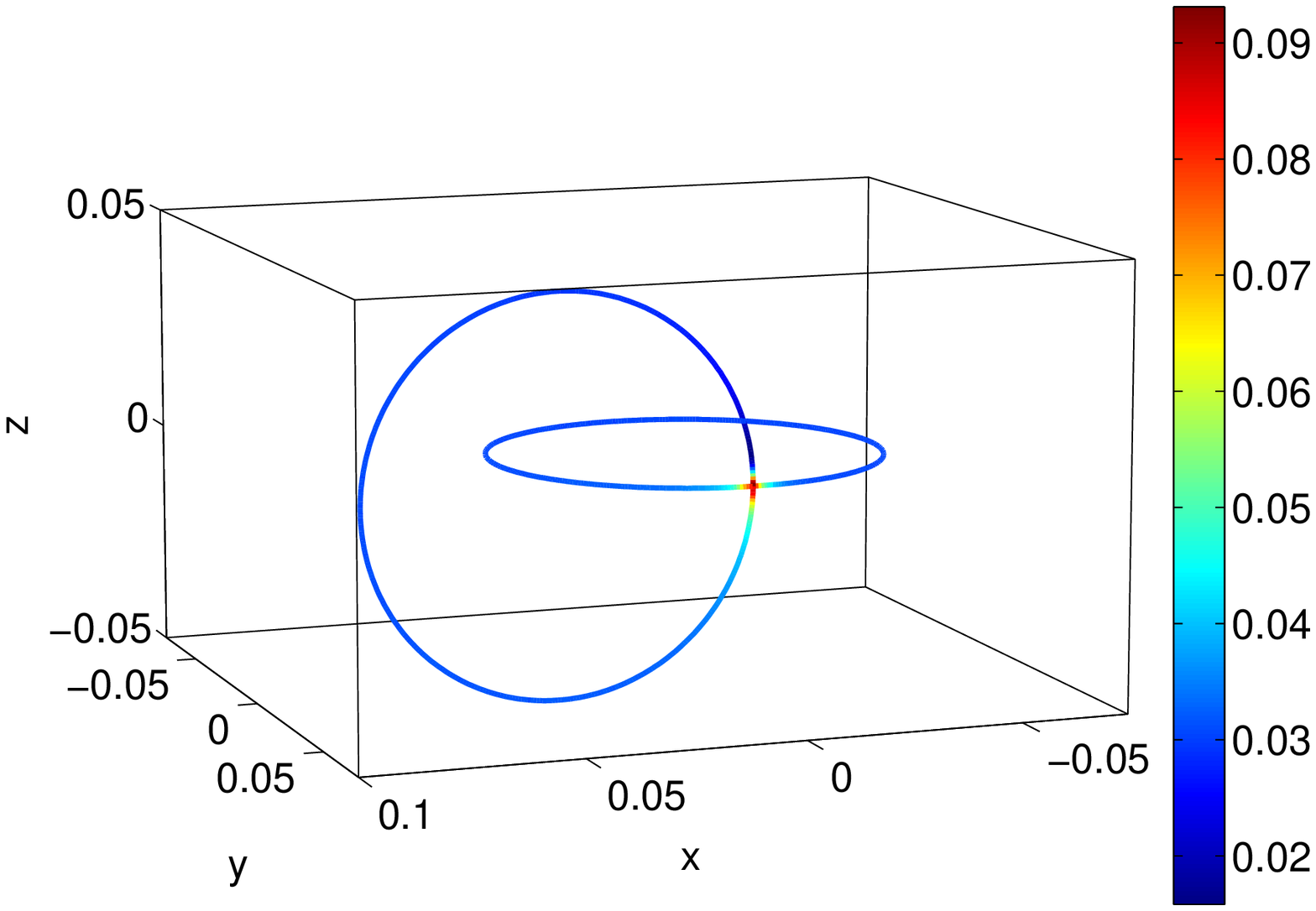}\\
\includegraphics[scale=.39]{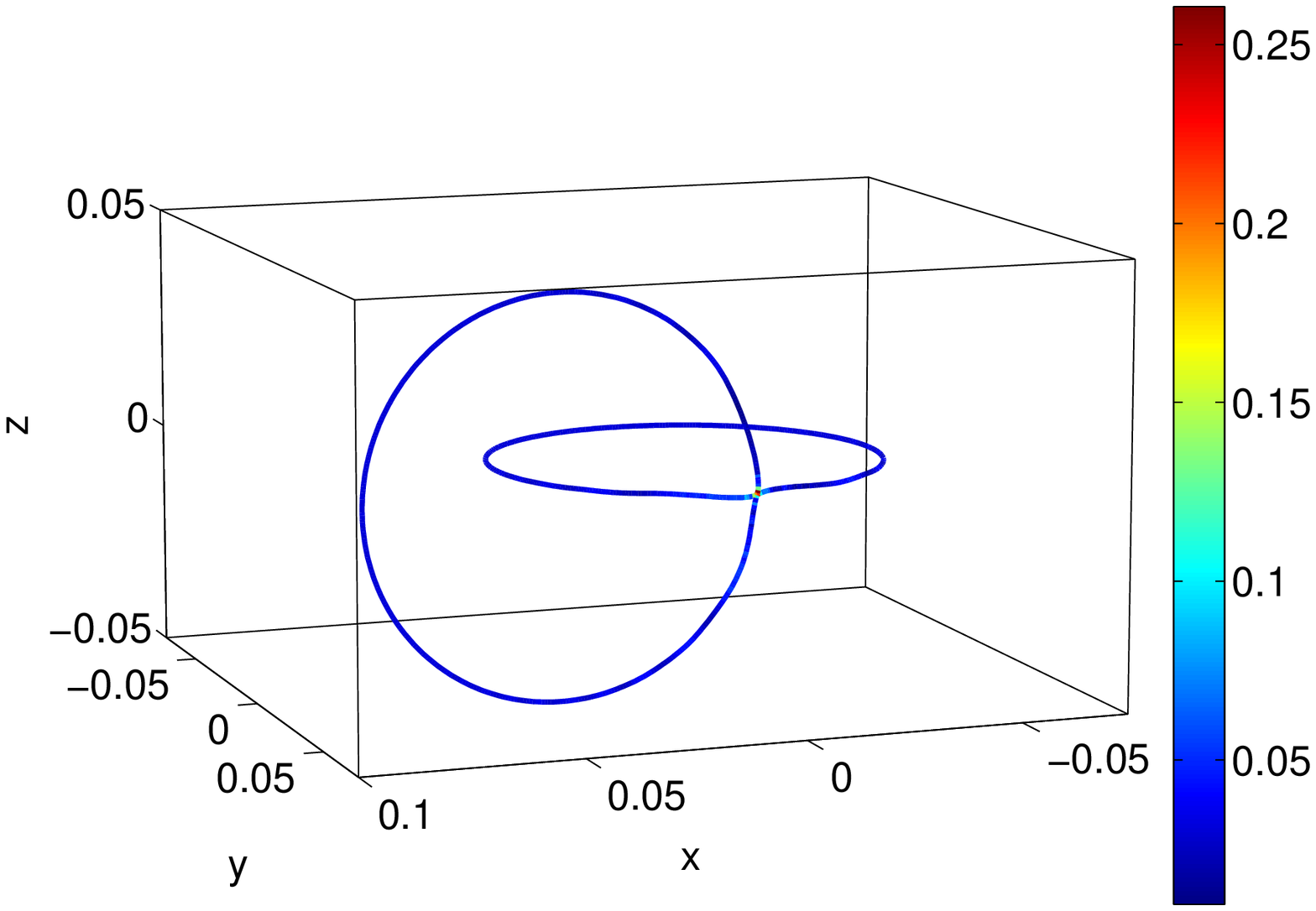}\\
\includegraphics[scale=.39]{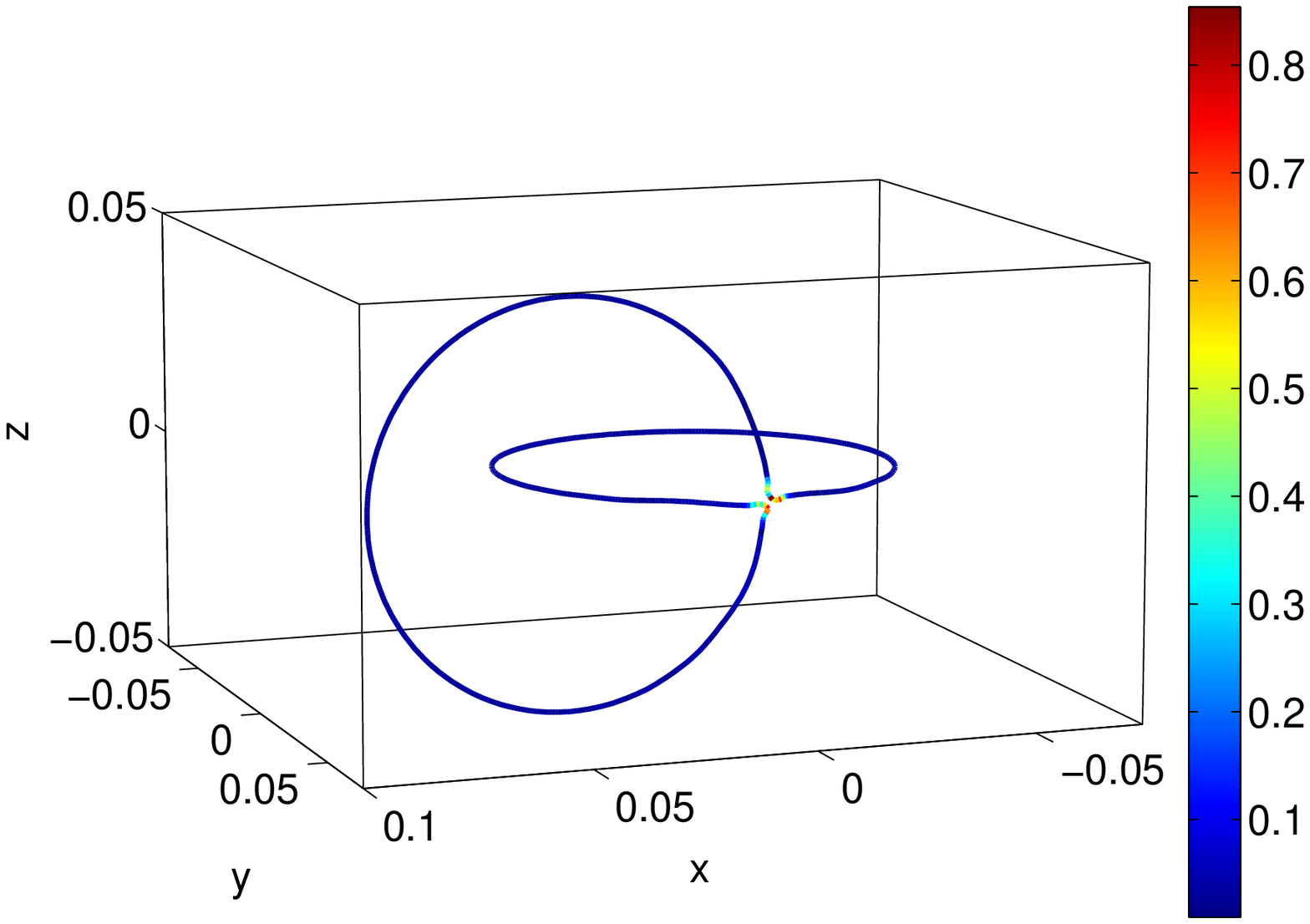} 
\end{center}
\caption{Reconnection of two vortex rings initially set perpendicular
to each other, computed with the Biot-Savart law:
Top: at time $t=0$;  middle: $t_0-t=0.001~\rm s$; 
bottom: $t-t_0=0.005~\rm s$. The vortex lines
are colour-coded to indicate the magnitude of the velocity (in $\rm cm/s$,
see legend on each figure). The box is for visualization only.
}
\label{fig:13}
\end{figure}

\newpage

\begin{figure}[h!!!]
\begin{center}
\includegraphics[scale=.55]{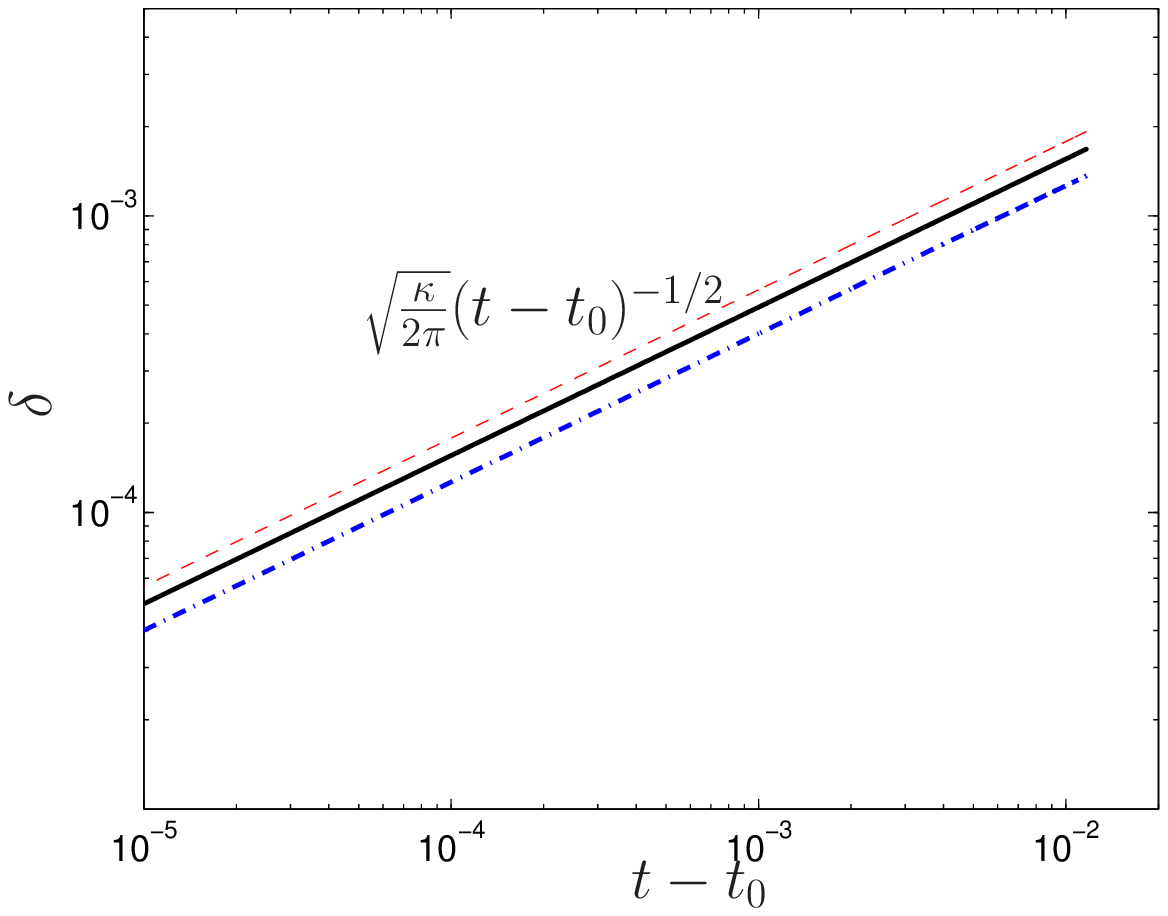}\\
\vspace{1cm}
\includegraphics[scale=.55]{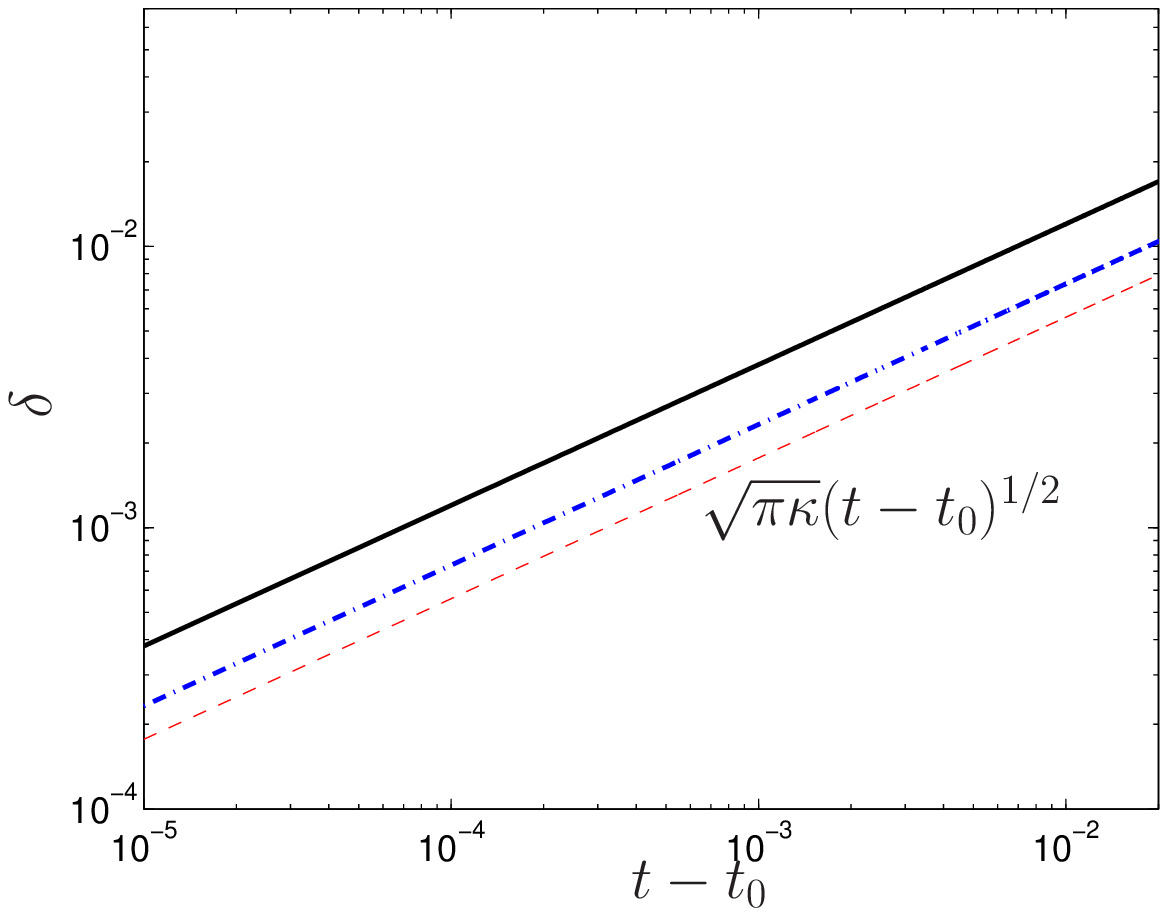}
\end{center}
\caption{ 
Minimum distance between the filaments $\delta(t)$ ($\rm cm$)
vs $(t_0-t)$ ($\rm s$)
before the reconnection (top) and vs $(t-t_0)$ after the reconnection
(bottom),
corresponding to the Biot-Savart evolution of two parallel
vortex rings shown in figure~(\ref{fig:11}) (solid black line) and
of two perpendicular vortex rings shown in figure~(\ref{fig:12})
(dot-dashed blue line).
$t_0$ is the time at which the reconnection takes place. The dashed
red line expresses $\delta(t)=\sqrt{\kappa (t_0-t)/(2 \pi)}$ found by
de Waele \& Aarts~\cite{WA1994} (top) and 
$\delta(t)=\sqrt{\pi \kappa (t-t_0)}$ (bottom) as a guide to the eye.
}
\label{fig:14}
\end{figure}

\end{document}